\documentstyle[12pt]{article}

\def\squarebox#1{\hbox to #1{\hfill\vbox to #1{\vfill}}}
\newcommand{\qed}{\hspace*{\fill}
            \vbox{\hrule\hbox{\vrule\squarebox{.667em}\vrule}\hrule}\smallskip}
 
\newtheorem{THEOREM}{Theorem}
\newenvironment{theorem}{\begin{THEOREM} \hspace{-.85em} {\bf :} \rm}%
                        {\end{THEOREM}}
\newtheorem{LEMMA}[THEOREM]{Lemma}
\newenvironment{lemma}{\begin{LEMMA} \hspace{-.85em} {\bf :} \rm}%
                      {\end{LEMMA}}
\newtheorem{COROLLARY}[THEOREM]{Corollary}
\newenvironment{corollary}{\begin{COROLLARY} \hspace{-.85em} {\bf :} \rm}%
                          {\end{COROLLARY}}
\newtheorem{PROPOSITION}[THEOREM]{Proposition}
\newenvironment{proposition}{\begin{PROPOSITION} \hspace{-.85em} {\bf :} \rm}%
                            {\end{PROPOSITION}}
\newenvironment{proof}{\noindent {\bf Proof:} \hspace{.677em}}%
                      {}
 
\newcommand{\DG}{D_{\sts G}}
\newcommand{\Dn}{D}
\newcommand{\sts}{\scriptscriptstyle}
\newcommand{\imp}{\supset}

\newcommand{\sat}{\models}
\newcommand\eqdef{\buildrel {\rm def}\over =}

\newcommand{\cM}{{\cal M}}

\newcommand{\subone}{_{\sts 1}}
\newcommand{\subtwo}{_{\sts 2}}
\newcommand{\subG}{_{\sts G}}

\newcommand{\pone}{p_{\sts 1}}
\newcommand{\ptwo}{p_{\sts 2}}

\newcommand{\bw}{\bigwedge\limits}
\newcommand{\C}{C}
\newcommand{\Cd}{C^\dm}
\newcommand{\Ce}{C^\epsilon }
\newcommand{\Cl}{C^{\scriptscriptstyle L}}

\newcommand{\Cts}{C^{\scriptscriptstyle {T}}}

\newcommand{\eps}{\epsilon}

\newcommand{\Ed}{E^\dm }
\newcommand{\Ee}{E^\epsilon }
\newcommand{\El}{E^{\scriptscriptstyle L}}

\newcommand{\Ets}{E^{\scriptscriptstyle {T}}}
\newcommand{\Etz}{E^{\scriptscriptstyle {T 0}}}

\newcommand{\K}[4]{{\cal K}^#1_#2(#3,#4)}

\newcommand{\dm}{\diamond}

\newcommand{\m}{{\rm {\bf m}}}
\newcommand{\gb}{\goodbreak}
\newcommand{\rz}{r_{\sts 0}}
\newcommand{\rone}{r_{\sts 1}}

\newcommand{\snt}{\hbox{{\it sent}}}
\newcommand{\tz}{t_{\sts 0}}
\newcommand{\tone}{t_{\sts 1}}
\newcommand{\ts}{{\sts {T}}}

\newcommand{\I}{{\cal I}}

\newcommand{ \D}{\diamondsuit}

\newcommand{\bbox}{\vrule height7pt width4pt depth1pt}
\newcommand{\ie}{i.e.,~}
\renewcommand{\qed}{\bbox}
\newcommand{\cf}{cf.~}
\newcommand{\comment}[1]{\marginpar{\scriptsize\raggedright #1}}
\newcommand{\markj}{\comment{}}
\renewcommand{\phi}{\varphi}
\newcommand{\stigma}{v}

\begin{document}
\begin{titlepage}
 
\title{\Large Knowledge and Common Knowledge \\
       in a Distributed Environment\thanks{This
                      is a revised and expanded
                      version of a paper with the same title
                      that first appeared in the {\em Proceedings
                      of the 3rd ACM Conference on Principles of
                      Distributed Computing, 1984}.  It is essentially
                      identical to the version that appears in
{\em Journal of the ACM\/} {\bf 37}:3, 1990, pp. 549--587.
                      The work of the second author was supported
                      in part by DARPA contract N00039-82-C-0250.}}
 
\author{
     \begin{tabular}{ccc}
         \large Joseph Y. Halpern & \hspace{.25in} & \large Yoram Moses \\
             & \\
         \small \vspace{-3pt} IBM Almaden Research Center &&
                \small Department of Applied Mathematics \\
         \small \vspace{-3pt} San Jose, CA 95120 & &
                 \small The Weizmann Institute of Science\\
                 & & \small Rehovot, 76100 ISRAEL
     \end{tabular}
        }
 
\date{ }
\maketitle
\thispagestyle{empty}

\noindent {\bf  Abstract:}\quad
Reasoning about knowledge seems to play a fundamental role in
distributed systems. Indeed, such reasoning is a central
part of the informal intuitive arguments used in the design of
distributed protocols. Communication in a distributed system can
be viewed as the act of transforming the system's state of knowledge.
This paper presents a general framework for formalizing and reasoning
about knowledge in distributed systems.
We argue that states of knowledge of groups of processors
are useful concepts for the design and analysis of distributed
protocols. In particular, {\em distributed knowledge\/}
corresponds to knowledge that is ``distributed''
among the members of the group, while {\em common knowledge\/}
corresponds to a fact being ``publicly known''.
The relationship between common knowledge and a variety of
desirable actions in a distributed system is illustrated.
Furthermore, it is shown that, formally speaking, in practical
systems common knowledge cannot be attained.
A number of weaker variants of common knowledge that are attainable
in many cases of interest are introduced and investigated.

\vfill
 
\end{titlepage}
\newcommand{\erase}[1]{}
 
\section{Introduction}
Distributed systems of computers are rapidly gaining
popularity in a wide variety of applications.
However, the distributed nature of control and information
in such systems makes the design and analysis of distributed
protocols and plans a complex task. In fact, at the current
time, these tasks are more an art than a science.
Basic foundations, general techniques, and a clear methodology are
\markj%
needed to improve our understanding and ability to
deal effectively with distributed systems.
 
\markj%
While the tasks that distributed systems are required
to perform are normally stated in terms of the global
behavior of the system,
the actions that a processor performs can depend only
on its local information.
Since the design of a distributed protocol involves determining the
behavior and interaction between individual processors in the system,
designers frequently find it useful to reason intuitively about
processors' ``states of knowledge'' at various points in the execution of
a protocol. For example, it is customary to argue that
``\hbox{$\ldots$} once the
sender receives the acknowledgement, it {\em knows} that the current
packet has been delivered; it can then safely discard the current
packet, and send the next packet\hbox{$\ldots$}''.
Ironically, however, formal descriptions of distributed protocols,
as well as actual proofs of their correctness or impossibility,
have traditionally avoided any explicit mention of knowledge.
Rather, the intuitive arguments about the state of
knowledge of components of the system are customarily
buried in combinatorial
proofs that are unintuitive and hard to follow.
 
The general concept of knowledge has received considerable attention in a
variety of fields, ranging from Philosophy \cite{Hi1} and
 Artificial Intelligence \cite{MSHI} and \cite{Mo},
to Game Theory \cite{Au} and Psychology \cite{ClM}.
The main purpose of this paper is to demonstrate the relevance of
reasoning about knowledge to distributed systems as well.
Our basic thesis is that explicitly
reasoning about the states of knowledge of the components of a
distributed system provides a more general and uniform setting that
\markj%
offers insight into the basic structure and limitations
of protocols in a given system.

As mentioned above, agents can only
base their actions on their local information. This knowledge,
in turn,   depends on the messages they receive and the
events they observe. Thus, there is a close
relationship between knowledge and action
in a distributed environment.
When we consider the task of performing coordinated actions
among a number of agents in a distributed environment, it
does not, in general, suffice to talk only about individual agents'
knowledge. Rather, we need to look at states of knowledge of
{\em groups} of agents (the group of all participating agents
is often the most relevant one to consider).
Attaining particular states of group knowledge is a prerequisite
for performing coordinated actions of various kinds.
 
In this work we define a hierarchy of states of group knowledge.
It is natural to think of communication in the system as the
act of improving the state of knowledge, in the sense of
``climbing up the hierarchy''.
The weakest state of knowledge we discuss is {\em distributed knowledge},
which corresponds to knowledge that is distributed among the members of
the group, without any individual agent necessarily having it.%
\footnote{In a previous version of this paper \cite{HM1}, what we
are now calling distributed knowledge was called implicit knowledge.
We have changed the name here to avoid conflict with the usage of the
phrase ``implicit knowledge'' in papers such as \cite{FH,Lev2}.}
The strongest state of knowledge in the hierarchy is
{\em common knowledge}, which roughly corresponds to ``public
knowledge''. We show that the execution of simultaneous actions
becomes common knowledge, and hence that such actions cannot be
performed if common knowledge cannot be attained.
Reaching agreement is an important example of a desirable
simultaneous action in a distributed environment.
A large part of the technical analysis in this paper is concerned
with the ability and cost of attaining common
knowledge in systems of various types.
It turns out that attaining common knowledge in distributed
environments is not a simple task. We show that when communication
is not guaranteed it is impossible to attain common knowledge.
This generalizes the impossibility of a solution to
the well-known {\em coordinated attack} problem \cite{Gray}.
A more careful analysis shows that common
knowledge can only be attained in systems that support
simultaneous coordinated actions.
It can be shown that such actions cannot be guaranteed or detected
in practical distributed systems.
It follows that common knowledge cannot be attained in many cases
of interest.
We then consider states of knowledge that correspond to eventually
coordinated actions and to coordinated actions that are guaranteed to
be performed within a bounded amount of time. These are essentially
weaker variants of common knowledge. However, whereas, strictly
speaking,  common knowledge may be difficult to attain in many
practical cases, these weaker states of knowledge are attainable in
cases of interest.

Another question that we consider is that of
when it is safe to assume that
certain facts are common knowledge, even when strictly speaking
they are not. For this purpose, we introduce the  concept
of {\em internal knowledge consistency}.
Roughly speaking, it is internally knowledge consistent to assume
that a certain state of knowledge holds at a given point, if nothing
the processors in the system will ever encounter will be inconsistent
with this assumption.
 
\erase{
When dealing with distributed systems,
reasoning about the state of knowledge of all processors
in the system, and how their knowledge changes over time,
turn out to be useful. To this end, we introduce a whole hierarchy
of states of knowledge that a group may be in.  Communication in a
distributed system can often be viewed as the act of ``climbing up the
hierarchy''.  For example, if processor~$\pone$ receives a
message $m$ from $\ptwo$, then we can say
$\pone$ knows $m$.  If $\pone$ acknowledges
receiving this message and $\ptwo$ receives the acknowledgement,
then $\ptwo$ knows that $\pone$ knows~$m$.  If $\ptwo$ acknowledges the
acknowledgement and this is received by $\pone$, then $\pone$ knows
that $\ptwo$ knows that $\pone$ knows $m$, and so on.
 
Of particular
interest to us here is the notion of {\em common knowledge}, which
is the highest level of knowledge in our hierarchy.
Common knowledge of a fact $\varphi$ arises when everyone in
the group knows $\varphi$, everyone knows that everyone knows
$\varphi$, and so on.
Common knowledge is a prerequisite for agreement, and
for simultaneous coordinated action.  We show that common
knowledge is not attainable in systems where communication is
not guaranteed, and as a corollary we show that
there is no solution to the well-known {\em coordinated attack}
problem.
 
A closer look at common knowledge shows that not
only is it not attainable in systems where communication is
not guaranteed, but it is also not attainable in systems
where communication is guaranteed, but there is some uncertainty
about message delivery time.
This observation leads us to take a closer look at common knowledge
and the interaction between knowledge and communication.
We introduce a number of weaker variants of common knowledge that
are attainable in some cases of interest.  For example,
{\em eventual common knowledge} is attainable in situations
where messages are guaranteed to arrive eventually, but there
is no upper bound on message delivery time, while {\em likely common
knowledge} is attainable if message delivery is likely.
By analyzing these weaker variants of common knowledge we can gain
a better understanding of the relationship between knowledge
and action in a distributed system.
}
 
The rest of the paper is organized as follows.
In the next section we look at the ``muddy children'' puzzle,
which illustrates some of the subtleties involved in reasoning about
knowledge in the context of a group of agents.
In Section~3 we introduce a hierarchy of states of knowledge
in which a group may be.
Section~4 focuses on the relationship between knowledge
and communication by looking at the coordinated attack problem.
In Section~5 we sketch a general definition of a distributed system,
and in Section~6 we discuss how knowledge can be ascribed to processors
in such systems so as to make statements such as ``agent 1
{\em knows} $\varphi$'' completely formal and precise.
Section~7
relates common knowledge to the coordinated attack problem.
\markj%
In Section~8, we show that, strictly speaking, common knowledge cannot
be attained in practical distributed systems.
Section~9 considers the implications of this observation
and in Section~10 we begin to reconsider the notion of common knowledge
in the light of these implications.
In  Sections~11 and~12, we consider a number of variants of
common knowledge that are attainable in many cases of interest and
discuss the relevance of these states of knowledge to the actions
that can be performed in a distributed system.
Section~\ref{internal} discusses the notion of
internal knowledge consistency, and Section~\ref{conc}
contains some concluding remarks.

\section{The muddy children puzzle}
 
A crucial aspect of distributed protocols is the fact that a number
of different processors cooperate in order to achieve a particular goal.
In such cases, since more than one agent is present,
an agent may have knowledge about other
agents' knowledge in addition to his knowledge about
the physical world.
This often requires care in distinguishing
subtle differences between seemingly similar states of knowledge.
A classical
example of this phenomenon is the muddy children puzzle --
a variant of the well known ``wise men'' or ``cheating wives'' puzzles.
The version given here is taken from \cite{Bar}:
 
\begin{quote}
 
\quad Imagine $n$ children playing together.
The mother of these children has told
them that if they get dirty there will be severe consequences.
So, of course,
each child wants to keep clean, but each would love to see the others get
dirty.  Now it happens during their play that some of
the children, say $k$
of them, get mud on their foreheads.  Each can see the mud on others but
not on his own forehead. So, of course, no one says a thing.  Along comes
the father, who says, ``At least one of you has mud on your head,'' thus
expressing a fact known to each of them before he spoke (if $k>1$).
The father then
asks the following question, over and over:  ``Can any of you
prove you have mud on your head?''  Assuming that all the
children are perceptive, intelligent,
truthful, and that they answer simultaneously, what will happen?
\end{quote}
 
The reader may want to think about the situation before reading
the rest of Barwise's discussion:
 
\begin{quote}
\quad There is a ``proof'' that the first $k-1$ times he asks
the question, they will all say ``no'' but then
the $k$th time the dirty children will answer ``yes.''
 
\quad The ``proof'' is by induction on $k$.  For $k=1$
the result is obvious:  the
dirty child sees that no one else is muddy, so he
must be the muddy one.  Let
us do $k=2$. So there are just two dirty children,
$a$ and $b$.  Each answers ``no''
the first time, because of the mud on the other.
But, when $b$ says ``no,'' $a$ realizes that he must
be muddy, for otherwise $b$
would have known the mud was on
his head and answered ``yes'' the first time.
Thus $a$ answers ``yes'' the
second time.  But $b$ goes through the same reasoning.
Now suppose $k=3$; so
there are three dirty children, $a, b, c$.
Child $a$ argues as follows.  Assume
I don't have mud on my head.  Then, by the $k=2$ case,
both $b$ and $c$ will
answer ``yes'' the second time.  When they don't,
he realizes that the assumption was false, that he is muddy,
and so will answer ``yes'' on the third question.
Similarly for $b$ and $c$.
[The general case is similar.]
\end{quote}
 
Let us denote the fact ``At least one child has a muddy forehead'' by
{\bf m}.
Notice that if $k>1$, \ie more than one child has a muddy forehead,
then every child can see at least one muddy forehead, and
the children initially all know {\bf m}.
Thus, it would seem, the father does not need to tell
the children that {\bf m} holds when $k>1$.
But this is false! In fact,
had the father not announced {\bf m},
the muddy children would never have been able
to conclude that their foreheads are muddy.
We now sketch a proof of this fact.
 
First of all, given that the children are intelligent and truthful,
a child
with a clean forehead will never answer ``yes'' to any of the father's
questions. Thus, if $k=0$, all of the children answer all of the
father's questions ``no''.
Assume inductively that if there are exactly~$k$ muddy children
and the father does not announce {\bf m}, then the children all
answer ``no'' to all of the father's questions. Note that, in particular,
when there are exactly~$k$ muddy foreheads, a child with a clean
forehead initially sees $k$ muddy foreheads and hears all of the father's
questions answered ``no''.
Now assume that there are exactly $k+1$ muddy children.
Let $q\ge 1$ and assume that all of the children answer ``no''
to the father's first $q-1$ questions.
We have argued above that a clean child will necessarily answer ``no''
to the father's $q^{\rm th}$ question. Next observe that
before answering the father's $q^{\rm th}$ question, a
muddy child has exactly the same information as a clean child has
at the corresponding point in the case of~$k$ muddy foreheads.
It follows that the muddy children must all answer ``no'' to the father's
$q^{\rm th}$ question, and we are done.  (A very similar
proof shows that if there are~$k$ muddy children and the
father does announce {\bf m}, his first~$k-1$ questions are
answered ``no''.)
 
So, by announcing something that the children all know, the father
somehow manages to give the children useful information!
How can this be? Exactly what {\em was} the role of the
father's statement?
In order to answer this question, we need to take a closer
look at knowledge
in the presence of more than one knower; this is the subject of the next
section.
 
\section{A hierarchy of states of knowledge}
 
In order to analyze the muddy children puzzle introduced in the previous
section, we need to consider states of knowledge of groups
of agents.
As we shall see in the sequel, reasoning about such states of knowledge
is crucial in the context of distributed systems as well.
In Section~6 we shall carefully define what it means for an
agent~$i$
to know a given fact~$\varphi$ (which we denote by $K_i\varphi$).
For now, however, we need knowledge to satisfy only two properties.
The first is that an agent's knowledge at a given time must depend
only on its local history: the information that it started out with
combined with the events it has observed since then.
Secondly, we require that only true things be known, or
more formally:
$$\quad K_i\varphi \imp \varphi\hbox{;}$$ \ie
if an agent $i$ knows $\varphi$, then $\varphi$ is true.
This property, which is occasionally
referred to as the {\em knowledge axiom},
is the main property that
philosophers customarily use to distinguish knowledge from belief
(\cf \cite{HM2}).
 
 Given a reasonable interpretation for what it means for an
agent to know a fact~$\varphi$, how does the notion of knowledge
generalize from an agent to a group?
In other words, what does it mean to say that a group~$G$ of
agents knows a fact $\varphi$? We believe that more than one
possibility is reasonable, with the appropriate choice depending
on the application:
 
\begin{itemize}
\item $\DG\varphi$ \quad (read ``the group $G$ has
{\em distributed knowledge} of $\varphi$''):
We say that knowledge of $\varphi$ is distributed in $G$ if
someone who knew everything that
each member of $G$ knows would know $\varphi$.
For instance, if one member of $G$ knows $\psi$ and another knows that
$\psi\imp \varphi$, the group $G$ may be said to have distributed
knowledge of $\varphi$.
\item $S\subG\varphi$ \quad (read ``{\em someone in~$G$
knows~$\varphi$}''):
We say that $S\subG\varphi$ holds iff some member of $G$ knows $\varphi$.
More formally, $$S\subG\varphi\equiv \bigvee_{i\epsilon G} K_i\varphi.$$
\item $E\subG\varphi$ \quad (read ``{\em everyone in
$G$ knows $\varphi$}''):
We say that $E\subG\varphi$ holds iff all members of $G$ know $\varphi$.
More formally, $$E\subG\varphi\equiv \bigwedge_{i\epsilon G}
K_i\varphi.$$
\item $E\subG^k\varphi$, for $k\ge 1$ \quad
 (read ``$\varphi$ is $E^k$-{\em knowledge} in $G$''):\quad
$E\subG^k\varphi$ is defined by
$$E\subG^1\varphi = E\subG \varphi,$$
$$E\subG^{k+1}\varphi = E\subG E\subG^k\varphi,\ {\rm for}\ k\ge 1.$$
$\varphi$ is said to be $E^k$-knowledge in~$G$ if ``everyone
in~$G$ knows that everyone in~$G$ knows
that $\ldots$ that everyone in~$G$ knows that $\varphi$ is
true'' holds, where the phrase ``everyone in~$G$ knows that''
appears in the sentence~$k$ times.
\item $C\subG\varphi$ \quad (read ``$\varphi$ is
{\em common knowledge} in $G$''):
The formula~$\varphi$ is said to be common knowledge in~$G$
if $\varphi$ is  $E\subG^k$-knowledge for all $k\ge 1$.
In other words,
$$C\subG\varphi\equiv E\subG \varphi
\wedge E\subG^2\varphi
\wedge\cdots\wedge E\subG^m\varphi\wedge\cdots$$
\end{itemize}
(We omit the subscript $G$ when the group $G$ is understood from context.)

Clearly, the notions of group knowledge introduced above
form a hierarchy, with
$$C\varphi\imp\cdots\imp E^{k+1}\varphi\imp\cdots\imp E\varphi\imp
S\varphi\imp \Dn\varphi\imp \varphi.$$
However, depending on the circumstances, these notions
might not be distinct.
For example, consider a model of parallel computation
in which a collection of
$n$ processors share a common memory.
If their knowledge is based on the contents of the common
memory, then we arrive
at a situation in which
$C\varphi\equiv E^k\varphi\equiv E\varphi\equiv S\varphi\equiv \Dn\varphi$.
By way of contrast, in a distributed system in which $n$ processors are
connected via some communication network and each one of them
has its own memory, the above hierarchy is strict.
Moreover, in such a system, every
two levels in the  hierarchy can be separated by an
actual task, in the sense
that there will be an action for which one level in the hierarchy will
suffice, but no lower level will. It is quite clear that this is the case
with $E\varphi\imp S\varphi\imp \Dn\varphi$, and, as we are about to show,
the ``muddy children'' puzzle is an example of a situation in which
$E^k \varphi$ suffices to perform a required action,
but $E^{k-1} \varphi$ does not.
In the next section we present the coordinated attack problem,
a problem for which $C\varphi$ suffices to perform a required action,
but for no $k$ does $E^k\varphi$ suffice.
 
Returning to the muddy children puzzle, let us
consider the state of the children's knowledge of {\bf m}: ``At least
one forehead is muddy''.
Before the father speaks,  $E^{k-1}${\bf m} holds,
and $E^k${\bf m} doesn't. To see this, consider the case $k=2$
and suppose that Alice and Bob are the only muddy children.
Clearly everyone sees at least one muddy child, so $E{\bf m}$ holds.
But the only muddy child that Alice sees is Bob, and, not knowing whether
she is muddy, Alice considers it possible that Bob
is the only muddy child.
Alice therefore considers it possible that Bob sees no muddy child.
Thus, although both Alice and Bob know {\bf m} (\ie $E\m$ holds),
Alice does not know that Bob knows~$\m$, and hence
$E^2{\bf m}$ does not hold.
A similar argument works for the general case.
We leave it to the reader to check that when there
are $k$ muddy children,
$E^k${\bf m} suffices to ensure that the muddy children will
be able to prove their dirtiness, whereas $E^{k-1}${\bf m} does not.
(For a more detailed analysis of this argument, and for a general
treatment of variants of the muddy children puzzle, see \cite{MDH}.)
 
Thus, the role of the father's statement was to improve the children's
state of knowledge of {\bf m} from $E^{k-1}${\bf m} to $E^k${\bf m}.
In fact, the children have {\em common knowledge} of {\bf m}
after the father announces that  {\bf m} holds.
Roughly speaking, the father's public announcement of
{\bf m} to the children as a group results in all the children
knowing {\bf m} and knowing that the father has publicly
announced {\bf m}.
Assuming that it is common knowledge that all of the children know
anything the father announces publicly, it is easy to conclude
that the father's announcement makes $\m$ common knowledge.
Once the father announces $\m$, all of the children
know both $\m$ and that
the father has
announced $\m$. Every child thus knows that all of the
children know $\m$ and know that the father publicly announced $\m$,
and so $E^2\m$ holds.
It is similarly possible to show that
once the father announces $\m$ then $E^k\m$ holds for all $k$,
\markj%
so $C\m$ holds (see Section~10 for further discussion).
Since, in particular, $E^k${\bf m} holds, the muddy
children can succeed in proving their dirtiness.
 
The vast majority of the communication in a distributed system can also
be viewed as
the act of improving the state of knowledge (in the
sense of ``climbing up a
hierarchy'') of certain facts. This is an elaboration of the view of
communication in a network as the act of ``sharing
knowledge''. Taking this
view, two notions come to mind. One is {\em fact discovery} --
the act of changing the state of knowledge of a fact $\varphi$
from being distributed knowledge to
levels of explicit knowledge (usually $S$-, $E$-, or $C$-knowledge),
and the other is {\em fact  publication} --
the act of changing the state of
knowledge of a fact that is  not common knowledge to common knowledge.
An example of fact discovery is the detection of
global properties of a system, such as deadlock. The
system initially has distributed knowledge of the deadlock,
and the detection algorithm improves this state to $S$-knowledge
(see \cite{ChL} for work related to fact discovery).
An example of fact publication is the introduction of a new
communication convention in a computer network. Here
the initiator(s) of the convention wish to make the new convention
common knowledge.
 
In the rest of the paper we devote a considerable amount of
attention to fact publication and common knowledge.
As we shall show, common knowledge is inherent in a variety of notions
of agreement, conventions, and coordinated action.
Furthermore, having common knowledge of a large number of
facts allows for more efficient communication.
Since these are goals frequently sought in distributed computing,
the problem of fact publication --- how to
attain common knowledge --- becomes crucial.
Common knowledge is also a basic notion in everyday communication
between people. For example, shaking hands to seal an agreement
signifies that the handshakers have common knowledge of the agreement.
Also, it can be argued \cite{ClM}
that when we use a definite reference such as
``the president'' in a sentence, we assume
common knowledge of who is being referred to.
 
In \cite{ClM}, Clark and Marshall present two basic
ways in which a group can come to have common knowledge of a fact.
One is by membership in a community, e.g., the meaning of a
red traffic light is
\markj%
common knowledge in the community of licensed drivers.
The other is by being copresent with the occurrence of the fact, e.g.,
the father's gathering the children and publicly announcing the
existence of muddy foreheads made that fact common knowledge. Notice that
if, instead, the father had
taken each child aside (without the other children noticing)
and told her or him about it privately, this information would
have been of
no help at all.

In the context of distributed systems, community
membership corresponds to information that the processors are
guaranteed to have by virtue of their presence in the system
(e.g., information that is ``inserted into'' the
processors  before they enter the system).
However, it is not obvious how to simulate copresence
or ``public'' announcements using message passing
in a distributed system.
As we shall see, there are serious problems and unexpected
subtleties involved in attempting to do so.
 
\section{The coordinated attack problem}
 
To get a flavor of the issues involved in attaining common knowledge
by simulating copresence in a distributed system,
consider the coordinated attack problem,
originally introduced by Gray \cite{Gray}:
 
\begin{quote}
 
Two divisions of an army are camped on two hilltops overlooking
a common valley. In the valley awaits the enemy.
It is clear that if both divisions attack the enemy
simultaneously they will win
the battle, whereas if only one division attacks it will be defeated.
The divisions do not initially have plans for launching
an attack on the enemy, and
the commanding general of the first division wishes to coordinate a
simultaneous  attack (at some time the next day).
Neither general will decide to
attack unless he is sure that the other will attack with him.
The generals can only communicate by means of a messenger.
Normally, it takes the messenger one hour to get from one
encampment to the other.  However, it is possible that he will
get lost in the dark or, worse yet, be captured by the enemy.
Fortunately, on this particular night, everything goes smoothly.
How long will it take them to coordinate an attack?
\end{quote}

We now show that despite the fact that everything goes smoothly, no
agreement can be reached and no general can decide to attack.
(This is, in a way, a folk theorem of operating systems theory; \cf
\cite{Gal,Gray,YC}.)
Suppose General $A$ sends a message to General $B$ saying
``Let's attack at dawn'', and the messenger delivers it
an hour later. General $A$ does not immediately know
whether the messenger
succeeded in delivering the message.
And because $B$ would not attack at dawn if the messenger is
captured and fails to deliver the message,
$A$ will not attack unless he
knows that the message was successfully delivered.
Consequently, $B$ sends the messenger back to $A$
with an acknowledgement.
Suppose the messenger delivers the acknowledgement
to $A$ an hour later.
Since $B$ knows that $A$ will not attack without knowing
that $B$ received the original message, he knows that $A$ will not attack
unless the acknowledgement is successfully delivered.
Thus, $B$ will not attack unless he knows
that the acknowledgement
has been successfully delivered.
However, for $B$ to know that the acknowledgement
has been successfully
delivered, $A$ must send the messenger back with
an acknowledgement to the acknowledgement \dots .
Similar arguments can be used to show that no fixed
finite number of acknowledgements, acknowledgements to acknowledgements,
etc.\ suffices for the generals to attack.
Note that in the discussion above the generals are essentially running a
{\em handshake} protocol (\cf \cite{Gray}). The above discussion shows that
for no $k$ does a $k$-round handshake
protocol guarantee that the generals be able to coordinate an attack.
 
In fact, we can use this intuition to actually prove that
the generals can never attack and be guaranteed that they are attacking
simultaneously.
We argue by induction on $d$ --- the number of
messages delivered by the time of the attack ---
that $d$ messages do not suffice.
Clearly, if no message is delivered, then $B$ will not know of
the intended attack, and a simultaneous attack is impossible.
For the inductive step, assume that $k$ messages do not suffice.
If $k+1$ messages suffice, then the sender of
the $(k+1)^{\rm st}$ message
attacks without knowing whether his last message arrived. Since
whenever one general attacks they both do, the intended
receiver of the $(k+1)^{\rm st}$ message must attack regardless of whether
the $(k+1)^{\rm st}$ message is delivered. Thus, the $(k+1)^{\rm st}$
message
is irrelevant, and $k$ messages suffice,
contradicting the inductive hypothesis.
 
After presenting a detailed proof of the fact that no
protocol the generals
can use will satisfy their requirements and allow them to
coordinate an attack, Yemini and Cohen in \cite{YC} make the
following remark:
 
\begin{quote}
\ldots Furthermore, proving protocols correct
(or impossible) is a difficult and cumbersome art in the absence of
proper formal tools to reason about protocols.
Such backward-induction argument as the one used in the impossibility
proof should require less space and become more convincing with a
proper set of tools.
\end{quote}
 
Yemini and Cohen's proof does not explicitly use reasoning
about knowledge,
but it uses a many-scenarios argument to show that if the
generals both attack in one scenario, then there is another
scenario in which one general will attack and the other will not.
The crucial point is that the actions that should be
taken depend not only on the actual state of affairs (in this case,
the messenger successfully delivering the messages), but also
(and in an acute way) on what other states of affairs the
generals consider {\em possible}.
Knowledge is just the dual of possibility, so
reasoning about knowledge
precisely captures the many-scenario argument in an intuitive way.
We feel that understanding the role knowledge plays
in problems such as coordinated attack
is a first step towards simplifying the task of designing and
proving the correctness of protocols.
 
A protocol for the coordinated attack problem, if one did exist,
would ensure  that when the
generals attack, they are guaranteed to be attacking simultaneously.
Thus, in a sense, an attacking general (say $A$)
would know that the other general (say $B$) is also attacking.
Furthermore, $A$ would know that $B$ similarly knows
that $A$ is attacking.
It is easy to extend this reasoning to show that when the generals
attack they have common knowledge of the attack.
However, each message that the messenger delivers can add
at most one level
of knowledge about the desired attack, and no more.
For example, when the message is first delivered to $B$,
$B$ knows
about $A$'s desire to coordinate an attack, but $A$ does not know
whether the message was delivered, and therefore $A$ does
not know that $B$
knows about the intended attack. And when the messenger
returns to $A$ with $B$'s
acknowledgement, $A$ knows that $B$ knows about
the intended attack,
but, not knowing whether the messenger delivered the acknowledgement,
$B$ does
not know that $A$ knows (that $B$ knows of the intended attack).
This in some sense explains  why the generals cannot
reach an agreement to attack using a finite number of messages.
We are about to formalize this intuition.
Indeed, we shall prove a more general result from which the
inability to achieve a guaranteed
coordinated attack  will follow as a corollary.
Namely, we prove that communication cannot be used to attain
common knowledge in a system in which communication is not guaranteed,
and formally relate a guaranteed coordinated attack to attaining
common knowledge.
Before we do so, we need to define some of the terms that we use
more precisely.

\section{A general model of a distributed system}
 
We now present a general model of a distributed environment.
Formally,
we model such an environment by a distributed
system, where the agents are taken to be processors and interaction
between agents is modeled by messages sent between the processors
over communication links. For the sake of generality and
applicability to problems involving synchronization in distributed
systems, our treatment will allow processors to have hardware clocks.
Readers not interested in such issues can safely ignore all reference
to clocks made throughout the paper.
 
We view a distributed system as a finite collection
$\{\pone,\ptwo,\ldots,p_n\}$ of two or more processors that
are connected by a communication network.
We assume an external source of ``real time'' that in general
is not directly observable by the processors.
The processors are state machines that possibly have {\em clocks},
where a clock is a monotone nondecreasing function of real time.
If a processor has a clock, then we assume that
its clock reading is part of
its state.  (This is in contrast to the approach taken by Neiger
and Toueg in \cite{NT}; the difference is purely a matter of taste.)
The processors communicate with each other by
sending messages along the links in the network.

A {\em run} $r$ of a distributed system is a description of
an execution
of the system, from time 0 until the end of the execution.
(We assume for simplicity that the system executes forever.
If it terminates after finite time, we can just assume that
it remains in the same state from then on.)
A {\em point} is a pair $(r,t)$ consisting of a run $r$
and a time $t \ge 0$.
We characterize the run $r$ by associating with
each point $(r,t)$ every processor $p_i$'s {\em local history}
at $(r,t)$, denoted $h(p_i,r,t)$.  Roughly speaking,
$h(p_i,r,t)$ consists of the sequence of
events that $p_i$ has observed up to time $t$ in run $r$.
We now formalize this notion.
\markj %
We assume that processor $p_i$ ``wakes up'' or joins the system in run
$r$ at some time $t_{init}(p_i,r)
\ge 0$.
\markj%
The processor's local
state when it wakes up is called its {\em initial state}.
The {\em initial configuration\/} of a run consists of the
initial state and the wake up time for each processor.
In systems with clocks, the {\em clock time function\/} $\tau$
describes processors' clock readings;
$\tau(p_i,r,t)$ is the
reading of $p_i$'s clock at the point $(r,t)$.  Thus, $\tau(p_i,r,t)$
is undefined for $t < t_{init}(p_i,r)$ and is a monotonic nondecreasing
function of $t$ for $t \ge t_{init}(p_i,r)$.
We say that
$r$ and $r'$ have the same clock readings if
$\tau(p_i,r,t) = \tau(p_i,r',t)$ for all processors $p_i$ and
all times $t$.  (If there are no clocks in the system, we say
for simplicity that all runs have the same clock readings.)
We take $h(p_i,r,t)$ to be empty if $t < t_{init}(p_i,r)$.
For $t \ge t_{init}(p_i,r)$, the history $h(p_i,r,t)$
consists of $p_i$'s initial state and the sequence of messages
$p_i$ has sent
and received up to, but not including, those sent or received at time
$t$ (in the order
they were sent/received).  We assume that this sequence of messages is
finite.  If $p_i$ has a clock, the messages are also
marked with the time
at which
they were sent or received (\ie with $\tau(p_i,r,t)$, if they were
sent or received at time $t$), and the history
includes the range of values that the clock has read up to
and including time $t$.
\markj %
If we consider randomized protocols, then $h(p_i,r,t)$ also includes
$p_i$'s random coin tosses.  For ease of exposition, we restrict
attention to deterministic protocols in this paper.
In a deterministic system with no external inputs and no failures,
a processor's internal state will be a function of its history.
Thus, the sequence of internal
states that a processor goes through can be recovered from its
history.
 
Corresponding to every distributed system,
given an appropriate set of assumptions about the
properties of the system and its possible interaction with
its environment, there is a natural set~$R$ of all possible
runs of the system.
We identify a distributed system with
such a set $R$ of its possible runs.
For ease of exposition, we sometimes slightly
abuse the language and talk about a point $(r,t)$
as being {\em a point of}~$R$ when $r\in R$.
\markj%
A run $r'$ is said to {\em extend\/} a point $(r,t)$ if $h(p_i,r,t') =
h(p_i,r',t')$ for all $t' \le t$ and all processors $p_i$.
Observe that~$r'$ extends $(r,t)$ iff~$r$ extends $(r',t)$.
 
Identifying a system with a set of runs is an important idea that
will play a crucial role in allowing us to make precise the
meaning of knowledge in a distributed system.
The relative behavior of clocks, the properties
of communication in
the system, and many other properties of the system,
are directly reflected in the properties of this set of runs.
Thus, for example, a system is synchronous exactly if in all possible runs
of the system the processors and the communication
medium work in synchronous phases.
A truly asynchronous system is one in which the set of runs allows
any message sent to be delayed an unbounded amount of time before
being delivered. (We discuss asynchrony in greater detail
\markj%
in Section~8.)  Clocks are guaranteed to be
synchronized to within a bound of~$\delta$ if they differ
by no more than~$\delta$ time units at all points in all runs of
the system.
If we view the set of runs as a probability space with some appropriate
measure, then we can also capture probabilistic properties of the
environment and formalize probabilistic protocols in this framework.
 
We shall often be interested in the set of runs generated by running
a particular {\em protocol}, under some assumptions on the communication
medium.  Intuitively, a protocol is a function specifying what
actions a processor takes (which in our case amounts to what
messages it sends) at any given point (after the processor wakes
\markj%
up) as a function of the processor's local state.
Since a processor's local state is determined by its history,
we simply define a protocol to be a deterministic
function specifying what messages the processor
should send at
any given instant, as a function of the processor's history.
Recall that $h(p_i,r,t)$, processor $p_i$'s history at the
point $(r,t)$, does not include messages sent or
received at time $t$, so a processor's actions at time $t$
according to a protocol depend only on messages received in
the past.
\markj %
As we mentioned above,
for ease of exposition we restrict attention to deterministic protocols
in this paper.
The definitions and results can be extended
to nondeterministic and
probabilistic protocols in a straightforward way.  A {\em joint protocol
for} $G$ is a tuple consisting of a protocol for every processor in~$G$.

\section{Ascribing knowledge to processors}
 
What does it mean to say that a processor {\em knows} a fact~$\varphi$?
In our opinion, there is no unique ``correct'' answer
to this question.
Different interpretations of knowledge in a distributed
system are appropriate for different applications.
For example, an interpretation by which a processor is said to
know~$\varphi$ only if~$\varphi$ appears explicitly in a designated part
of the processor's storage (its ``database'') seems interesting
for certain applications.
In other contexts we may be interested in saying that a processor
knows~$\varphi$ if the processor could deduce~$\varphi$ from the
information available to it%
.  %
In this section we give precise definitions of interpretations
of knowledge in a distributed system.
 
We assume the existence of an underlying logical language
of formulas for representing {\em ground} facts about the system.
A ground fact is a fact about the state of the system
that does not explicitly involve processors' knowledge.
For example, {\em ``the value of register $x$ is~0''},
or {\em ``processor~$p_i$ sent the message~$m$ to
processor~$p_j$''}, are ground facts.

We extend the original language of ground formulas to a language that
is closed under operators for knowledge,
distributed knowledge, everyone knows, and common knowledge
(so that for every formula $\varphi$, processor $p_i$, and subset $G$ of
the processors, $K_i \varphi$,
$\DG\varphi$, $E\subG\varphi$, and $C\subG\varphi$
are formulas), and under Boolean connectives.
\markj%
(In Section~11 we consider additional operators.)
 
We now describe one of the most natural ways of ascribing knowledge
to processors in a distributed system, which we call {\em view-based\/}
knowledge interpretations.
At every point each processor is assigned a {\em view}; we say that
two points are {\em indistinguishable\/} to the processor if it
has the same view in both.
A processor is then said to {\em know\/} a fact at a given point
exactly if the fact holds at all of the points that the processor
cannot distinguish from the given one.
Roughly speaking, a processor knows all of the facts that
(information theoretically) follow from
its view at the current point.%
\footnote{In a previous version of this paper \cite{HM1}, view-based
knowledge interpretations were called {\em state-based}
interpretations. Particular view-based knowledge interpretations were
first suggested to us independently by Cynthia Dwork and by
Stan Rosenschein.  Since the appearance of \cite{HM1}, most authors who
considered knowledge in distributed systems have focussed on
view-based interpretations; \cf \cite{ChM,DM,FI1,HF,LR,MT,PR,RK}
and \cite{Hal1} for an overview.
(See \cite{FH,Mos} for examples of interpretations of knowledge
that are not view based.)
The approach taken to defining knowledge in view-based systems
is closely related to the possible-worlds approach taken by Hintikka
\cite{Hi1}.
For us the ``possible worlds'' are the points in the system; the
``agents'' are the processors.  A processor in one world
(\ie point) considers another world possible if it has the same
view in both.}
 
More formally,
a {\em view function}~$\stigma$ for a system~$R$ assigns
to every processor at any given point of~$R$ a view
from some set $\Sigma$ of {\em views} (the structure
of~$\Sigma$ is not relevant at this point);
\ie $\stigma(p_i,r,t)\in \Sigma$ for each
processor~$p_i$ and point $(r,t)$ of~$R$.
Given that a processor's history captures all of the events
in the system that a processor may possibly observe,
we require the processor's view at any given point to
be a function of its history at that point.
In other words, whenever
$h(p_i,r,t)=h(p_i,r',t')$, it must also be the case that
$\stigma(p_i,r,t)=\stigma(p_i,r',t')$.

A {\em view-based knowledge interpretation} $\I$ is a triple $(R,\pi,
\stigma)$, consisting of a
 set of runs~$R$, an assignment~$\pi$ which associates with every
point in $R$ a truth assignment to the ground facts (so that for
every point $(r,t)$ in $R$ and every ground fact $P$, we have
$\pi(r,t)(P) \in \{{\bf true, false}\}$),
and a view
function~$\stigma$ for~$R$.
A triple $(\I,r,t)$, where~$\I$ is a
knowledge interpretation and $(r,t)$ is a point of~$R$,
is called a {\em knowledge point}.
Formulas are said to be true or false of knowledge points.
Let $\I=(R,\pi,\stigma)$. We can now define the truth of
a formula~$\varphi$
at a knowledge point $(\I,r,t)$, denoted $(\I,r,t)\sat\varphi$
(and also occasionally read {\em ``$\varphi$ holds at $(\I,r,t)$''},
or just {\em ``$\varphi$ holds at $(r,t)$''}, if the interpretation
$\I$ is clear from context),
by induction on the structure of formulas:
\begin{itemize}
\item[(a)] If~$P$ is a ground formula then
$(\I,r,t)\sat P$ iff $\pi(r,t)(P) = {\bf true}$.
\item[(b)] $(\I,r,t)\sat\neg\psi$ iff $(\I,r,t)\not\sat\psi$.
\item[(c)] $(\I,r,t)\sat\psi\subone\wedge\psi\subtwo$ iff
$(\I,r,t)\sat\psi\subone$ and $(\I,r,t)\sat\psi\subtwo$.
\item[(d)] $(\I,r,t)\sat K_i\psi$ iff
$(\I,r',t')\sat\psi$ for all $(r',t')$ in $R$ satisfying
$\stigma(p_i,r,t)=\stigma(p_i,r',t')$.
\end{itemize}

Part (a) says the truth value of ground facts is defined by $\pi$.
Parts~(b) and~(c)
state that negations and conjunctions have their classical meaning.
Part~(d) captures the fact a processor~$p_i$'s knowledge
at a point $(r,t)$ is completely determined by its view
$\stigma(p_i,r,t)$. The processor does not know~$\varphi$ in
a given view exactly if there is a point (in~$R$) at which
the processor has that same view, and~$\varphi$ does
not hold.
The definitions of when $E\subG\varphi$ and $C\subG\varphi$
hold at a knowledge point follow directly from the definition of
$E\subG$ and $C\subG$ in Section~3:
 
\begin{itemize}
\item[(e)] $(\I,r,t)\sat E\subG\psi$ iff
$(\I,r,t)\sat K_i\psi$ for all  $p_i\in G$.
\item[(f)] $(\I,r,t)\sat C\subG\psi$ iff
$(\I,r,t)\sat E\subG^k\psi$ for all $k>0$.
\end{itemize}

Let us consider when a group~$G$ of processors has distributed knowledge
of a fact.
Intuitively, a group's distributed knowledge is the combined
knowledge of all of its members. For example, we could imagine
considering the group as being able to distinguish two points if
one (or more) of its members can distinguish them.
The set of points indistinguishable by~$G$ from the current one is
then the intersection of the sets of points indistinguishable by
the individual members of the group.
We can therefore define when a group~$G$ has distributed knowledge of
a fact~$\varphi$
as follows:
\begin{itemize}
\item[(g)] $(\I,r,t)\sat \DG\psi$ iff
$(\I,r',t')\sat\psi$ for all  $(r',t')$ in $R$ satisfying
$\stigma(p_i,r,t)=\stigma(p_i,r',t')$ for all $p_i\in G$.
\end{itemize}
 
\noindent Notice that indeed under this definition, if one member of~$G$
knows~$\varphi$ while another member knows that $\varphi\imp\psi$,
then the members of $G$ have distributed knowledge of $\psi$.
The definition of distributed knowledge given above
is in a precise sense a direct
generalization of the definition of individual processors' knowledge in
clause (d) above. We can define the {\em joint view
assigned by $\stigma$ to~$G$\/} to be
$$\stigma(G,r,t)\ \ \eqdef\ \
\{\langle p_i,\stigma(p_i,r,t)\rangle\,:\,p_i\in G\}.$$
It is easy to check that
$(\I,r,t)\sat \DG\psi$ iff
$(\I,r',t')\sat\psi$ for all  $(r',t')$ in $R$ satisfying
$\stigma(G,r,t)=\stigma(G,r',t')$.
Thus, we can identify the distributed knowledge of a group $G$ with
the knowledge of an agent whose view is the group's joint view.%
\footnote{The knowledge ascribed to a set of processes by
Chandy and Misra in \cite{ChM}
essentially corresponds to the distributed knowledge of that set,
as defined here. See also \cite{PR,RK}.}
Note that the knowledge distributed in a group of size one
coincides with its unique member's knowledge.
 
View-based interpretations will prove to be a
useful way of ascribing knowledge to processors for the purpose of
the design and analysis of distributed protocols. We now discuss some
of the basic properties of knowledge in view-based interpretations.
Fix a system $R$ and a view function $\stigma$.  We can construct a graph
corresponding to $R$ and $v$ by taking the nodes of the graph to be
all the points of $R$, and joining two nodes $(r,t)$ and $(r',t')$ by an
edge labelled $p_i$ if $v(p_i,r,t) = v(p_i,r',t')$; \ie
if $p_i$ has the same view at both points.
Our definition of knowledge under a view-based
 interpretation
immediately implies that $K_i\varphi$ holds at a given point $(r,t)$
if and only if $\varphi$ holds at all points $(r',t')$
 that share an edge labeled~$p_i$ with $(r,t)$.
Define a point $(r',t')$ in this graph to be
$G$-{\em reachable from $(r,t)$ in~$k$ steps (with respect to the view
 function $v$)} if there exist points $(\rz,\tz)$, $(\rone,\tone)$,
\ldots,
$(r_k,t_k)$ such that $(r,t)=(\rz,\tz)$, $(r',t')=(r_k,t_k)$, and
for every $i<k$ there is a processor $p_{j_i}\in G$
such that $(r_i,t_i)$ and $(r_{i+{\sts 1}},t_{i+{\sts 1}})$ are
joined
by an edge labeled $p_{j_i}$.  It follows
that $E\subG\varphi$ holds at $(r,t)$ under this view-based
interpretation exactly
if~$\varphi$ holds at all points $G$-reachable from $(r,t)$ in~1 step.
An easy induction on~$k$ shows that $E\subG^k\varphi$ holds exactly
if~$\varphi$ holds at all points $G$-reachable in~$k$ steps.
Consequently, it is easy to see that $C\subG\varphi$ holds at
a point $(r,t)$ if and only if~$\varphi$ holds at all points that are
$G$-reachable from $(r,t)$ in a finite number of steps.
 In the particular case that~$G$ is the set of all processors,
then $C\subG\varphi$ holds at $(r,t)$ exactly if~$\varphi$
holds at all points in the same connected component of the graph
as $(r,t)$.
 
The way distributed knowledge is represented
in this graph is also instructive: $\DG\varphi$ holds at a
given point $(r,t)$ iff~$\varphi$ holds at all points $(r',t')$
such that for each $p_i \in G$, there is an
edge between $(r,t)$ and $(r',t')$ labelled by $p_i$.
Thus, for distributed knowledge the set of points we need to consider
is the intersection of the sets of points we consider when determining
what facts each individual processor knows.
 
By describing the various notions of knowledge in the view-based
case via this graph, it becomes easier to investigate their properties.
In fact, this graph is very closely related to {\em Kripke
structures}, a well known standard way of modeling modal logics.
In fact, drawing on the theory of modal logics, we can immediately
see that the definition of knowledge in view-based interpretations
agrees with the well-known modal logic S5 (\cf \cite{HM2}).
A modal operator~$M$ is said to {\em have the properties of S5} if it
satisfies the following axioms and rule of inference:
\begin{itemize}
\item[A1.] The {\em knowledge axiom}: $M\varphi\imp\varphi$,
\item[A2.] The {\em consequence closure axiom}:
              $M\varphi\wedge M(\varphi\imp\psi)\,\imp\,M\psi$,
\item[A3.] The {\em positive introspection axiom}:
              $M\varphi\imp MM\varphi$,
\item[A4.] The {\em negative introspection axiom}:
        $\neg M\varphi\imp M\neg M\varphi$, and
\item[R1.] The  {\em rule of necessitation}:
              From $\varphi$ infer $M\varphi$.
\end{itemize}
 
Given a knowledge interpretation~$\I$ for a system~$R$, a fact $\psi$
is said to be {\em valid in the system} if it holds at all knowledge
points $(\I,r,t)$ for points $(r,t)$ of $R$.
In our context the rule~R1 means that whenever~$\varphi$ is
valid in the system, so is $M\varphi$.
 
We can now show:
 
\begin{proposition}
Under view-based knowledge interpretations,
the operators $K_i$, $\DG$, and $C\subG$ all have
the properties of~S5.
\end{proposition}
 
The proof is a consequence of the fact that the definitions of
these notions are based on equivalence relations (over points):
The relation of processor~$p_i$'s having the same view
at two points, the relation of all processors in~$G$ having the same
joint views at both points, and the relation of being
reachable via a path consisting solely of edges
labeled by members of~$G$ in the graph corresponding to the view, are
all equivalence
relations. The proof of this proposition can be found in \cite{HM2}.

In addition to having the properties of S5,
common knowledge has two
additional useful properties under view-based interpretations:
 
\begin{itemize}
\item[C1.] The  {\em fixed point axiom}:
           $C\subG\varphi\equiv E\subG (\varphi\wedge C\subG\varphi)$, and
\item[C2.] The {\em induction rule}:  From
 $\varphi\imp E\subG(\varphi \land \psi)
 $ infer $\varphi\imp C\subG\psi$.
\end{itemize}
 
The fixed point axiom essentially characterizes $C\subG \varphi$
as the solution of a fixed point equation (in fact, it is the
\markj%
greatest solution; we discuss this in more detail in Section~11
and Appendix A).  This property of common knowledge is crucial
in many of our proofs.

Intuitively, the induction rule says
that if~$\varphi$ is ``public'' and implies $\psi$, so that
whenever~$\varphi$ holds then everybody knows~$\varphi \land \psi$,
then whenever $\phi$ holds, $\psi$ is common knowledge.
We call it the ``induction rule'' because it is closely
related to the notion of induction in arithmetic:
Using the fact that $\varphi\imp E\subG(\varphi \land \psi)$
   is valid in the system, we can
prove by induction on~$k$ that $\varphi\imp E\subG^k(\varphi \land \psi)$
is also valid in the system, for all~$k > 0$. It then follows
that $\varphi\imp C\subG\psi$ is valid in the system.
Roughly speaking, this proof traces our line of reasoning when
we argued that
the children in the muddy children puzzle attain common knowledge
of the father's statement.
We can get an important special case of the Induction Rule by
taking $\psi$ to be $\phi$.  Since $E\subG(\varphi \land \phi)$
is equivalent to $E\subG\varphi$, we get that from $\phi \imp E\subG
\phi$ we can infer $\phi \imp C\subG \phi$.
 
A very important instance of view-based knowledge interpretations,
\markj%
that will  be used extensively from Section~11 on, is called the
{\em complete-history interpretation}.
Under this interpretation
we have $\stigma(p_i,r,t)\eqdef h(p_i,r,t)$.
That is, the processor's complete history is taken to be
the view on which the processor's knowledge is based.
(In a previous version
of this paper \cite{HM1},
this was called the {\em total view\/} interpretation.)
The complete-history interpretation makes the finest possible distinctions
among histories. Thus, in a precise sense, it provides the
processors with at least as much
knowledge about the ground formulas as any
other view-based interpretation.
This is one of the reasons why the
complete-history interpretation
is particularly well suited for proving possibility and impossibility
of achieving certain goals in distributed systems, and for the
design and analysis of distributed protocols (\cf \cite{ChM,DM,MT}).
 
Notice that view-based knowledge interpretations ascribe knowledge
to a processor without the processor necessarily being
``aware'' of this knowledge, and without the processor
needing to perform any particular computation in order to
obtain such knowledge.
Interestingly, even if
 the view function~$\stigma$ does not distinguish
between possibilities at all, that is, if there is a single
view~$\Lambda$ such that $\stigma(p_i,r,t)=\Lambda$
for all~$p_i$, $r$, and~$t$,
the processors are still ascribed quite a bit of knowledge:
every fact that is true at all points of the system is common knowledge
among all the processors under this view-based interpretation
(and in fact under
all view-based interpretations).
Note that the hierarchy of Section~3 collapses
under this interpretation, with $\Dn\varphi\equiv E\varphi\equiv C\varphi$.
This interpretation makes the coarsest possible
distinctions among histories; at the other extreme we have the
complete-history interpretation, which makes the finest possible
distinctions among histories.
 
Another reasonable view-based interpretation is one in which
\markj%
$\stigma(p_i,r,t)$ is defined to be~$p_i$'s local state at
$(r,t)$.  (Recall that processors are state machines, and are thus
assumed to be in some local state at every point).
This is the choice made in \cite{FI1,Ros,RK}.
Under this interpretation, a processor might ``forget'' facts that it
knows. In particular, if a processor can
arrive at a given state by two different message histories, then,
once in that state, the processor's knowledge cannot distinguish between
these two ``possible pasts''. In the complete-history interpretation,
a processor's
view encodes all of the processor's previous states,
and therefore
processors do not forget what they know; if a processor
knows~$\varphi$ at
a knowledge point $(\I,r,t)$, then at all knowledge points $(\I,r,t')$
with $t'>t$ the processor will know that it once knew~$\varphi$.
Thus, while there may be temporary facts such as ``it is 3 on my clock''
which a processor will not know at 4 o'clock, it will know at 4~o'clock
that it previously knew that it was 3 o'clock.
 
Other view-based interpretations that may be of interest are
ones in which a processor's view is identified with
the contents of its memory, or with the position of its program counter
(see \cite{KT} for a closer look at some of these view-based
interpretations).   The precise view-based interpretation
we choose will vary from application to application.  For proving
lower bounds we frequently use the complete-history interpretation
since, in general, if processors cannot perform an action with
the knowledge they have in the complete-history interpretation, they
cannot perform it at all.  On the other hand, if we can show that
very little information is required to perform a given action, this
may suggest an efficient protocol for performing it.
 
\markj%
Although view-based knowledge interpretations are natural and
useful in many applications, they do not cover all reasonable
possibilities of ascribing knowledge to processors in a distributed
system. For example, as we have commented above, view-based
knowledge interpretations ascribe knowledge to processors
in a fashion that is independent of the processor's computational power.
To the extent that we intend processors' knowledge to closely
correspond to the actions they can perform, it often becomes crucial
to define knowledge in a way that depends on the processors'
computational powers (\cf \cite{MT,Mos}).
In most of the paper we deal exclusively with view-based
knowledge interpretations. However, in order to be able to
prove stronger negative
results about the attainability of certain states of knowledge,
we now give a general definition of knowledge interpretations, which
we believe covers
 all reasonable cases.

Intuitively, we want to allow any interpretation that satisfies the
two properties discussed in Section~3: (1) that a processor's knowledge
be a function of its history and (2) that only true things be known
(so that the axiom $K_i \varphi \imp \varphi$ is valid).
We capture the first property through the notion of an epistemic
interpretation.
An {\em epistemic interpretation\/} $\I$ is a function
assigning to every processor $p_i$ at any given point $(r,t)$,
a set $\K{\I}{i}{r}{t}$ of facts in the extended
language that $p_i$ is said to ``believe''.
$\K{\I}{i}{r}{t}$ is required to be a function of $p_i$'s
history at $(r,t)$.
Thus, if
 $h(p_i,r,t)=h(p_i,r',t')$, then
$\K{\I}{i}{r}{t} = \K{\I}{i}{r'}{t'}$.
 
Given an epistemic interpretation $\I$, we now specify when
a formula~$\varphi$ of the extended language holds at a point $(r,t)$
(denoted $(\I,r,t)\sat \varphi$).
As before, if $\varphi$ is a ground fact, we say that $(\I,r,t) \sat
\varphi$ iff $\pi(r,t)(\varphi) = {\bf true}$, while if
$\varphi$ is a conjunction or a negation, then
its truth is defined based on the truth of its subformulas in the
obvious way.
If~$\varphi$ is of the form $K_i\psi$, then
$(\I,r,t)\sat K_i\psi$ iff $\psi\in \K{\I}{i}{r}{t}$.
In this case we say that $p_i$ {\em believes\/} $\psi$.
The formula $E\subG\psi$ is identified with the
conjunction $\bw_{p_i\in G}K_i\psi$,
so that $(\I,r,t) \sat E\subG \psi$ iff $(\I,r,t) \sat K_i \psi$
for all $p_i \in G$.
If~$\varphi$ is of the form $C\subG\psi$,
then $(\I,r,t)\sat C\subG\psi$
iff $(\I,r,t) \sat E\subG (\psi \land C\subG\psi)$.
Thus, common knowledge is defined so that the fixed point axiom
\markj%
holds, rather than as an infinite conjunction.
Although this definition seems circular, it is not.   In order
to determine if $(\I,r,t) \sat C\subG\psi$, we first
\markj%
check if $(\I,r,t) \sat K_i(\psi \land
C\subG\psi)$ for all $p_i \in G$.  The
latter fact can be determined by considering the sets
$\K{\I}{i}{r}{t}$.
Finally, to handle distributed knowledge, we need to add a set
$\K{\I}{G}{r}{t}$ of formulas to every point $(r,t)$ for each set of
processors~$G$, analogous to the sets $\K{\I}{i}{r}{t}$ for individual
processors. We define $(\I,r,t)\sat \DG\varphi$ if
$\varphi\in\K{\I}{G}{r}{t}$. The sets $\K{\I}{G}{r}{t}$ must be a function
of~$G$'s joint history at $(r,t)$.  We may want to put some restrictions
on the sets $\K{\I}{G}{r}{t}$.  For example, we may require that
if $i\in G$ and $\varphi\in\K{\I}{i}{r}{t}$ then
$\varphi\in\K{\I}{G}{r}{t}$ (which implies that $K_i\varphi\imp \DG\varphi$
is valid).  Since we do not consider distributed knowledge in
interpretations that are not view based, we do not pursue the matter
any further here.

The knowledge axiom $K_i \varphi \imp \varphi$ is not necessarily
valid in epistemic interpretations.  Indeed, that is why
we have interpreted $K_i \varphi$ as ``processor $i$ believes $\varphi$''
in epistemic interpretations, since the knowledge axiom
is the key property that is taken
to distinguish knowledge from belief.  A processor's beliefs may be
false, although a processor cannot be said to {\em know\/} $\varphi$
if $\varphi$ is in fact false.
Given an epistemic interpretation~$\I$ and a set of runs~$R$, we say
that~$\I$ is a {\em knowledge interpretation for}~$R$ if
for all processors~$p_i$, times~$t$, runs $r\in R$ and formulas~$\varphi$
in the extended language, it is the case that
whenever $(\I,r,t)\sat K_i\varphi$ holds,  $(\I,r,t)\sat \varphi$
also holds.  Thus,
an epistemic interpretation for~$R$ is a knowledge interpretation for~$R$
exactly if it makes the knowledge axiom valid in $R$.
Notice that the view-based knowledge interpretations
defined above are in particular knowledge interpretations.

A trivial consequence of our definitions above is:
 
\begin{lemma}
\label{Lemma1}
 Let $\I$ be a knowledge interpretation for~$R$ and let $(r,t)$
be a point of~$R$.
The following are equivalent for a nonempty subset $G$ of processors:
\begin{enumerate}
\item
$(\I,r,t)\sat C\subG\varphi$
\item
\markj%
$(\I,r,t)\sat K_i(\varphi \land
C\subG\varphi)$ for all processors $p_i \in G$
\item
$(\I,r,t) \sat K_i(\varphi \land C\subG\varphi)$ for
some processor $p_i \in G$.
\end{enumerate}
\end{lemma}

This lemma shows that common knowledge requires simultaneity in a very
strong sense: When a new fact becomes common knowledge  in a group~$G$,
the local histories of all of the members of~$G$ must change
{\em simultaneously} to reflect the event of the fact's becoming common
knowledge.   This point is perhaps best understood if we think of time
as ranging over the natural numbers.  Given a knowledge interpretation
$\I$, suppose that  common knowledge does not hold at the point $(r,t)$ but
does hold at the point $(r,t+1)$, so that $(\I,r,t) \sat \neg C\subG\varphi$
and $(\I,r,t+1) \sat C\subG\varphi$.  Then it must be the case that
the local histories of all processors in $G$ changed between times $t$
and $t+1$.  To see this, note that by Lemma~\ref{Lemma1} we have
$(\I,r,t+1) \sat K_i (\varphi \land C\subG\varphi)$ for all
$p_i \in G$.  Suppose $p_i \in G$ has the same local history in
$(r,t)$ and $(r,t+1)$.  Then by our assumption that a processor's
knowledge depends only on its local history, we have that
$(\I,r,t) \sat K_i (\varphi \land C\subG\varphi)$.  Now by
Lemma~\ref{Lemma1} again, we have $(\I,r,t) \sat C\subG\varphi$,
contradicting our original assumption.
 
\markj%
We close this section with another trivial observation that follows
easily from Lemma~\ref{Lemma1}.
\begin{lemma}
\label{Lemma2}
Let $\I$ be a knowledge interpretation for $R$, let $r$ and $r'$ be runs
in $R$, and let $p_i$ be a processor in $G$.
If $h(p_i,r,t) = h(p_i,r',t')$ then $(\I,r,t) \models
C\subG \varphi$ iff $(\I,r',t') \models C\subG \varphi$.
\end{lemma}
\begin{proof}
Given that $p_i\in G$, we have by Lemma~\ref{Lemma1} that
$(\I,r,t) \models C\subG \varphi$ iff $(\I,r,t) \models
K_i(\varphi \land C\subG \varphi)$. Since $h(p_i,r,t) = h(p_i,r',t')$,
this holds iff $(\I,r',t') \models K_i(\varphi \land C\subG \varphi)$.
Again by Lemma~\ref{Lemma1} this is true iff
$(\I,r',t') \models C\subG \varphi$, and we are done.
\qed
\end{proof}
 
\section{Coordinated attack revisited}
 
Now that we have the basic terminology with which to define
distributed systems and knowledge in distributed systems,
we can relate the ability to perform a coordinated attack to
the attainment of common knowledge of particular facts.
This in turn will motivate an investigation of the attainability
of common knowledge in systems of various types.
 
We formalize the coordinated attack problem as follows:
We consider the generals as processors
and their messengers as communication links between them.
The generals are assumed to each behave according to
some predetermined deterministic protocol;
\ie a general's actions
(what messages it sends and whether it attacks) at a given point
are a deterministic function of his history and the time
on his clock.
In particular, we assume that the generals
are following a joint protocol $(P_A, P_B)$, where~$A$ follows~$P_A$
and~$B$ follows~$P_B$.
We can thus identify the generals with a distributed system~$R$,
consisting of all possible runs of $(P_A, P_B)$.
According to the description of the coordinated attack problem in
Section 4, the divisions do not initially
have plans to attack.
Formally, this means that the joint protocol the generals are following
has the property that in the absence of any successful communication
neither general will attack.  Thus, in any run of $R$ where no messages
are delivered, the generals do not attack.
 
We can now show that attacking requires attaining common knowledge
of the attack:
 
\begin{proposition}
\label{Proposition5}
Any correct protocol for the coordinated attack problem
has the property that whenever the generals attack,
it is common knowledge that they are attacking.
\end{proposition}
 
\begin{proof}
Let $(P_A, P_B)$ be a correct (joint) protocol for the coordinated attack
problem,  with~$R$
being the corresponding system.
Consider a ground language consisting of a single fact
$\psi\eqdef$ ``{\em both generals are attacking}'',
let~$\pi(r,t)$ assign a truth value to this formula in the obvious way
at each point $(r,t)$,
and let~$\I$ be the corresponding complete-history interpretation.
Assume that the generals attack at the point $({\hat r},{\hat t})$ of~$R$.
We show that $(\I,{\hat r},{\hat t})\sat C\psi$.  Our first step
is to show that $\psi\imp E\psi$ is valid in the system~$R$.
Assume that $(r,t)$ is an arbitrary
point of~$R$.  If $(\I,r,t) \sat  \neg \psi$, then we
trivially have $(\I,r,t) \sat \psi \imp E\psi$.  If $(\I,r,t) \sat \psi$,
then both generals attack at $(r,t)$.  Suppose that $(r',t')$ is
a point of $R$ in which $A$ has the same local history as in $(r,t)$.
Since $A$ is executing a
deterministic protocol and $A$ attacks in $(r,t)$, $A$ must also attack
in $(r',t')$.
Furthermore, given that the protocol is a correct protocol for coordinated
attack, if $A$ attacks in $(r',t')$, then so does $B$, and hence
$(\I,r',t')\sat\psi$. It follows that $(\I,r,t)\sat K_{\sts A}\psi$;
similarly we obtain $(\I,r,t) \sat K_{\sts B} \psi$.  Thus $(\I,r,t)
\sat E\psi$, and again we have $(\I,r,t) \sat \psi \imp E\psi$.
We have now shown that
$\psi\imp E\psi$ is valid in $R$.  By the induction rule it follows
that $\psi\imp C\psi$
is also valid in $R$.  Since
$(\I,{\hat r},{\hat t})\sat\psi$,
we have that $(\I,{\hat r},{\hat t})\sat C\psi$ and we are done.
\qed
\end{proof}

Proposition~\ref{Proposition5} shows that common knowledge
is a prerequisite for coordinated attack.
Unfortunately, common knowledge is not always
attainable, as we show in the next section.  Indeed,
it is the unattainability of common knowledge that
is the fundamental
reason why the generals cannot coordinate an attack.
 
\section{Attaining common knowledge}
\label{AttSection}
 
\markj%
Following the coordinated attack example, we first consider
systems in which communication is not guaranteed.
Intuitively, communication is not guaranteed in a system
if messages might fail to be delivered in an
arbitrary fashion, independent of any other event in the system.
Completely formalizing this intuition seems to be rather
cumbersome (\cf \cite{HF}), and we do not attempt to do so here.
For our purposes, a weak condition, which must be satisfied by any
reasonable definition of the notion of communication not being guaranteed,
will suffice. Roughly speaking, we take communication not being
guaranteed to correspond to two conditions.  The first says that it
is always possible that from some point on no messages will be received.
The second says that if processor $p_i$ does not get any information to
the contrary (by receiving some message), then $p_i$
considers it possible
that none of its messages were received.
 
Formally, given a system~$R$, we say that
{\em communication in~$R$ is not guaranteed} if the following two
conditions hold:
 
\begin{itemize}
\item[NG1.]
For all runs $r$ and times $t$, there exists a run $r'$ extending $(r,t)$
such that $r$ and $r'$ have the same initial configuration and the
same clock readings, and
no messages are received in $r'$ at or after time $t$.
\item[NG2.]
If in run $r$ processor $p_i$ does not receive any messages in the
interval $(t',t)$, then there is a run
$r'$ extending $(r,t')$ such that $r$ and $r'$ have the same initial
configuration and the same
clock readings,
$h(p_i,r,t'') = h(p_i,r',t'')$ for all $t'' \le t$, and
no processor $p_j \ne p_i$ receives a message in $r'$ in
the interval $[t',t)$.
\end{itemize}

\noindent
Note that the requirement that $r$ and $r'$ have the same initial
configuration already follows from the fact that $r'$ extends $(r,t)$ if
all the processors have woken up by time $t$ in run $r$.  In particular,
if we restricted attention to
systems where all processors were up at time 0, we would not require
this condition.
 
\markj%
We can now show that in a system in which communication is not
guaranteed, common knowledge is not attainable.
\begin{theorem}
\label{Lemma3}
Let $R$ be a system
in which communication
is not guaranteed,
  let $\I$ be a knowledge interpretation for $R$,
and let $\vert G \vert \ge 2$.
Let~$r$ be a run of~$R$, and let $r^-$ be a run of~$R$
with the same initial configuration and the same clock readings as $r$,
such that no messages are received in $r^-$ up to time~$t$.
Then for all formulas~$\varphi$ it
is the case that $(\I,r,t)\sat C\subG\varphi$ iff $(\I,r^-,t)\sat
C\subG\varphi$.
\end{theorem}
 
\begin{proof}
Fix~$\varphi$.  Without loss of generality, we can assume $\pone, \ptwo \in
G$.   Let $d(r)$ be the number of
\markj%
messages received in~$r$ up to (but not including)
time~$t$.  We show by induction on $k$ that if $d(r) = k$, then
$(\I,r,t)\sat C\subG\varphi$ iff $(\I,r^-,t)\sat C\subG\varphi$.
We assume that all the runs mentioned in the remainder of the
proof have the same initial configuration and the same clock
readings as $r$.
First assume that $d(r)=0$. Thus no messages are received in~$r$
up to time~$t$.
Since $r$ and $r^-$ have the same initial configuration and clock
readings, it follows that
$h(\pone,r,t)=h(\pone,r^-,t)$.
By Lemma~\ref{Lemma2} we have $(\I,r^-,t)\sat C\subG\varphi$
iff $(\I,r,t)\sat C\subG\varphi$, as desired.
 
Assume inductively that the claim
\markj%
holds for all runs $r'\in R$
with $d(r')=k$, and assume that $d(r)=k+1$.
Let $t'< t$ be the latest time at which a message is received in $r$
before time $t$. Let $p_j$ be a processor that receives a message
at time $t'$ in $r$. Let $p_i$ be a processor in $G$
such
that $p_i \ne p_j$ (such a $p_i$ exists since $\vert G \vert \ge 2$).
\markj%
From property NG2 in the
definition of communication not being guaranteed, it follows that there
is a
run $r'\in R$ extending $(r,t')$
such that $h(p_i,r,t'') = h(p_i,r',t'')$ for all $t'' \le t$ and
all processors $p_k\ne p_i$ receive no messages in $r'$ in the interval
$[t',t)$.  By construction, $d(r') \le k$, so
by the inductive hypothesis we have that
$(\I,r^-,t)\sat C\subG\varphi$ iff $(\I,r',t)\sat C\subG\varphi$.
Since $h(p_i,r,t) = h(p_i,r',t)$, by Lemma~\ref{Lemma2}
we have that $(\I,r',t)\sat C\subG\varphi$ iff $(\I,r,t) \sat
C\subG\varphi$.  Thus
$(\I,r^-,t)\sat C\subG\varphi$ iff $(\I,r,t)\sat C\subG\varphi$.
This completes the proof of the inductive step.
\qed
\end{proof}

\markj%
\markj%
Note
that Theorem~\ref{Lemma3} does not say that no fact
can become common knowledge in a system where communication is not
guaranteed.
In a system where communication is not
guaranteed but there is a global clock to which all processors
have access, then at 5 o'clock it becomes common knowledge
that it is 5 o'clock.%
\markj%
\footnote{We remark that the possible presence of some sort of
global clock
is essentially all that stops us from saying that
no fact can become common knowledge if it was not already common
knowledge at the beginning of a run.  See Proposition~\ref{new} in
Appendix B and the discussion before it for conditions under which it
is the case that no fact can become common knowledge which was not
initially common knowledge.}
However, the theorem does say that nothing can become common knowledge
unless it is also common knowledge in the absence of communication.
This is a basic property of systems with unreliable communication, and
it allows us to prove the impossibility of coordinated
attack.

\begin{corollary}
\label{Proposition5.1}
Any correct protocol for the coordinated attack
problem guarantees that neither party ever attacks (!).
\end{corollary}
 
\begin{proof}
Recall that communication between the generals is not
\markj%
guaranteed (\ie it satisfies conditions NG1 and NG2 above),
and we assume that in the absence of any successful
communication neither general will attack. Thus, if we take
$\psi$ to be
``{\em both generals are attacking}'',
then $C\psi$ does not hold at any point
in a run in which no messages are received (since $\psi$ does not hold
at any point of that run).
Theorem~\ref{Lemma3} implies that the generals will never
\markj%
attain common knowledge of $\psi$ in any run, and hence by
Proposition~\ref{Proposition5} the generals will never attack.  \qed
\end{proof}
 
It is often suggested that for any action for which $C\varphi$ suffices,
there is a~$k$ such that $E^k\varphi$ suffices, as is the case in the
muddy children puzzle.  The coordinated attack problem shows that this
is false.
The generals can
attain $E^k\varphi$ of many facts~$\varphi$
for an arbitrarily large~$k$
(for example, if the first $k$ messages are delivered).
However, simultaneous coordinated attack requires
{\em common knowledge\/} (as is shown in Proposition~\ref{Proposition5});
nothing less will do.
 
The requirement of simultaneous attack in the
coordinated attack problem is a very strong one.
It seems that real life generals do not need a protocol that
guarantees such a strong condition, and can probably make do
with one that guarantees a non-simultaneous attack.
We may want to consider weakening this requirement in order to
get something that is achievable.
In Section~11 we use a variant of the argument used in
\markj%
Corollary~\ref{Proposition5.1} to show that
no protocol can even guarantee that if one
party attacks then the other will {\em eventually} attack!
On the other hand, a protocol that guarantees that if one
party attacks, then with high probability the other will attack
{\em is\/} achievable, under appropriate probabilistic assumptions
about message delivery.  The details of such a protocol are
straightforward and left to the reader.
 
We can prove a result similar to Theorem~\ref{Lemma3}
even if communication is guaranteed, as long as there is no bound
on message delivery times.
A system $R$ is said to be a system with
{\em unbounded message delivery times}
if condition NG2 of communication not guaranteed holds, and in addition
we have:
\begin{itemize}
\item[NG$1'$.]
For all runs $r$ and all times $t$, $u$, with $t \le u$,
there exists a run $r'$ extending $(r,t)$ such that $r'$ has the same
initial configuration and the same clock readings
as $r$, and
no messages are received in $r'$ in the interval $[t,u]$.
\end{itemize}
 
\noindent
Asynchronous systems are often defined to be
systems with unbounded message delivery times (for example, in
\cite{FLP}).
Intuitively, condition NG$1'$
says that it is always possible for no
messages to be received for arbitrarily long periods of time, whereas
condition NG1
says that it is always possible for no messages at all to be received
from some time on.
In some sense, we can view NG1 as the limit case of
NG$1'$.
Notice that both systems where communication is not guaranteed and
systems with unbounded message delivery times satisfy condition
NG2.
The proof of
Theorem~\ref{Lemma3} made use only of NG2, not NG1, so
we immediately get

\begin{theorem}
Let $R$ be a system
with unbounded message delivery times,
  let $\I$ be a knowledge interpretation for $R$,
and let $\vert G \vert \ge 2$.
Let~$r$ be a run of~$R$, and let $r^-$ be a run of~$R$
with the same initial configuration and the same clock readings
as $r$,
such that no messages are received in $r^-$ up to time~$t$.
Then for all formulas~$\varphi$ it
is the case that $(\I,r,t)\sat C\subG\varphi$ iff $(\I,r^-,t)\sat
C\subG\varphi$.
\qed
\end{theorem}

The previous results show that,
in a strong sense, common knowledge is not attainable in  a system
in which communication is not guaranteed or, for that matter,
in a system in which
communication {\em is}
guaranteed, but there is no bound on the message delivery times.
However, even when all messages {\em are} guaranteed
to be delivered within a fixed time bound,
common knowledge can be elusive.
To see this, consider a system consisting of two
processors, R2 and D2, connected
by a communication link.
\markj%
Moreover, (it is common knowledge that)
communication is guaranteed.
But there is some uncertainty in message delivery times.
For simplicity, let us assume
that any message sent from R2 to D2
reaches D2 either immediately or after exactly $\epsilon$ seconds;
furthermore, assume that this fact is common knowledge.
Now suppose that
at time $t_S$, R2 sends D2 a message $m$ that does not contain a
timestamp, \ie does not mention $t_S$ in any way.
The message $m$ is received by D2 at time $t_D$.
Let $\snt(m)$ be the fact ``{\sl the message $m$ has been sent}''.
D2 doesn't know $\snt(m)$ initially. How does $\{{\rm R2, D2}\}$'s
state of knowledge of $\snt(m)$ change with time?
 
At time $t_D$, D2 knows $\snt(m)$. Because it might have
taken~$\eps$ time units
for $m$ to be delivered, R2 cannot be sure that D2 knows $\snt(m)$ before
$t_S+\eps$. Thus, $K_RK_D\snt(m)$ holds at time $t_S+\eps$
{\em and no earlier}.
D2 knows that R2 will not know that D2
knows $\snt(m)$ before $t_S+\eps$.
Because for all D2 knows $m$ may have been delivered
immediately (in which
case $t_S=t_D$), D2 does not know that R2 knows that D2
knows $\snt(m)$ before
$t_D+\eps$. Since $t_D$ might be equal to $t_S+\eps$,
R2 must wait until $t_S+2\eps$ before he knows that $t_D+\eps$
has passed. Thus, $K_RK_DK_RK_D\snt(m)$ holds at time $t_S+2\eps$
but no earlier.
This line of reasoning can be continued indefinitely, and
an easy proof by induction shows that before time $t_S+k\eps$, the
formula $(K_RK_D)^k\snt(m)$ does not hold, while at $t_S+k\eps$
it does hold.
Thus, it ``costs'' $\eps$ time units to acquire every level of
``R2 knows that D2 knows''.
Recall that $C\snt(m)$ implies $(K_RK_D)^k\snt(m)$ for every~$k$.
It follows that $C\snt(m)$ will never be attained!

We can capture this situation using our formal model as follows.
Let $MIN = \lfloor t_S/\eps \rfloor$, and consider
the system with a countable set of runs
$\{ r_i , r'_i:\,
\mbox{$i$ an integer with $i\ge -MIN$} \}$. If $i \ge -MIN$,
then in run $r_{i}$, R2 sends the message $m$ at time $t_S + i\eps$ and
D2 receives it at the same time.  In run $r_{i}'$, R2 again
sends the message $m$ at time $t_S + i\eps$, but D2 receives
it at time $t_S + (i+1)\eps$.  (Note our choice of $MIN$ guarantees
that all messages are sent at time greater than or equal to 0.)
If we assume that in fact the
message in the example
took $\eps$ time to arrive, then the run $r_0'$ describes
the true situation.  However, it is easy to see that at all
times $t$, R2 cannot distinguish runs $r_{i}$ and $r_{i}'$
(in that its local state is the same at the corresponding points
in the two runs, assuming that only message $m$ is sent), while
D2 cannot distinguish $r_{i}$ and $r_{i-1}'$ (provided $i-1 \ge -MIN$).

Our discussion of knowledge in a distributed system is motivated
by the fact that we can view processors' actions as being based
on their knowledge.
Consider an {\em eager\/} epistemic interpretation $\I$
under which R2 believes $C\snt(m)$ as soon as it sends the message
$m$, while D2
believes $C\snt(m)$ as soon as it receives $m$.
Clearly, $\I$ is not a knowledge interpretation, because it is not
knowledge consistent (R2 might believe that D2
knows $\snt(m)$, when in fact D2 does not).
However, once D2 receives $m$, which happens at most
$\eps$ time units after R2 starts believing
$C\snt(m)$,
it is easy to see that $C\snt(m)$ does indeed hold!
\markj%
In a sense, Lemma~\ref{Lemma1} says that
attaining common knowledge requires a certain kind of ``natural birth'';
it is not possible to attain it consistently unless simultaneity
is attainable.
But if one is willing to give up knowledge consistency
(\ie abandon the $K_i\varphi\imp \varphi$ axiom)
for short intervals of time, something very similar to common knowledge
can be attained.
 
The period of up to $\eps$ time units during which
R2 and D2's
``knowledge'' might be inconsistent might have many
negative consequences. If the processors need to act based on whether
$C\snt(m)$ holds during that interval, they might not
act in an appropriately coordinated way.
This is a familiar problem in the context of distributed
database systems.
There, committing a transaction roughly corresponds to entering
into an agreement that the transaction has taken place in the database.
However, in general, different sites of the database commit
transactions at different times (although usually all within a
small time interval).
When a new transaction is being committed
there is a ``window of vulnerability'' during which
different sites might reflect inconsistent histories of the database.
However, once all sites commit  the transaction, the history of the
database that the sites reflect becomes consistent
(at least as far as the particular transaction is concerned).
\markj%
In Section~13 we return to the question of when an ``almost knowledge
consistent'' version of
common knowledge can be safely used ``as if it were'' common knowledge.
 
Returning to the R2-D2 example, note that it is the uncertainty
in relative message delivery time
that makes it impossible to attain common knowledge, and
not the fact that communication is not instantaneous.
If it were common knowledge that messages took {\em exactly\/}
$\eps$ time units to arrive, then $\snt(m)$ would be common knowledge
at time $t_S + \eps$ (and the system would consist only of run
$r_1$).
 
Another way of removing the uncertainty is by having a common (global) clock
in the system.
Suppose that there is such a clock.
Consider what would happen if R2 sends D2 the following message $m'$:
 
\begin{quote}
``This message is being sent at time $t_S$; $m$.''
\end{quote}
 
\noindent
Since there is a global clock and it is guaranteed that
every message sent by R2 is delivered within $\eps$ time units,
the fact that R2 sent $m'$ to D2
would again become common knowledge at time $t_S+\eps$!
In this case, the system would consist of two runs, $r_0$ and
$r_1$.  At time $t_S+\eps$, D2 would know which of the two
was actually the case, although R2 would not (although D2
could tell him by sending a message).

It seems that common knowledge is attainable in
the latter two cases due to
the possibility of simultaneously making the transition from
not having common knowledge to having common knowledge
(at time $t_S+\eps$).
The impossibility of doing so in the first case was the driving
force behind the extra cost in time incurred in attaining each additional
level of knowledge.

Lemma~\ref{Lemma1} already implies that when $C\varphi$ first
holds all processors must come to believe $C\varphi$ simultaneously.
In particular, this means that all of the processors' histories must
change simultaneously. However, strictly speaking,
practical systems cannot guarantee absolute simultaneity.
In particular, we claim that essentially all practical
distributed systems have some inherent temporal uncertainty.
 There is always some uncertainty about
the precise instant at which each processor starts functioning, and
about exactly how much time each message takes to be delivered.
In Appendix~B we give a precise formulation of the notion of
{\em temporal imprecision}, which captures these
properties, and
 use methods derived from \cite{DHS} and \cite{HMM} to prove the
following result:
\begin{theorem}
\label{PracProp}
Let $R$ be a system with temporal imprecision,
let $\I$ be a knowledge interpretation for~$R$, and let $|G| \ge 2$.
Then for all runs $r\in R$, times~$t$, and formulas~$\varphi$
it is the case that $(\I,r,t)\sat C\subG\varphi$ iff
   $(\I,r,0)\sat C\subG\varphi$.
\end{theorem}

Since practical systems turn out to have temporal imprecision,
Theorem~\ref{PracProp} implies that, strictly speaking,
common knowledge cannot be attained in practical distributed systems!
In such systems, we have the following situation:
a fact~$\varphi$ can be known to a processor without being
common knowledge, or it can be common knowledge (in which case that
processor also knows~$\varphi$),
but due to (possibly negligible) imperfections in the system's
state of synchronization and its communication medium, there
is no way of getting from the first situation to the second!
Note that if there is a global clock, then there cannot
be any temporal imprecision.  Thus, it is consistent with
Theorem~\ref{PracProp} that common knowledge is attainable in
a system with a global clock.
 
Observe that we can now show that, formally speaking,
even people cannot attain common knowledge of any new fact!
Consider the father publicly announcing $\m$ to the children in
the muddy children puzzle.
Even if we assume that it is common knowledge
that the children all hear whatever the father says and understand it,
there remains some uncertainty as to exactly when each child comes
to know (or comprehend) the father's statement.
Thus, it is easy to see that the children do not immediately have common
knowledge of the father's announcement. Furthermore, for similar reasons
the father's statement can never become common knowledge.

\section{A paradox?}
 
There is a close correspondence between agreements,
coordinated actions, and common knowledge.
We have argued that in a precise sense,
reaching agreements and coordinating actions in a
distributed system requires attaining common knowledge
of certain facts.
However, in the previous section we showed
that common knowledge {\em cannot be attained}
in practical distributed systems!
We are faced with a seemingly paradoxical situation on
two accounts. First of all, these results are in contradiction with
practical experience, in which operations such as reaching agreement and
coordinating actions are routinely performed in many
actual distributed systems.
It certainly seems as if these actions are performed in such
systems without the designers having to worry about common knowledge
(and despite the fact that we have proved that common knowledge
is unattainable!).
Secondly, these results seem to contradict our intuitive feeling that
common knowledge {\em is} attained in many actual situations;
for example, by the children in the muddy children puzzle.
 
Where is the catch? How can we explain this apparent discrepancy
between our formal treatment and practical experience?
What is the right way to interpret our negative results
from the previous section? Is there indeed a paradox here?
Or perhaps we are using a wrong or useless definition of common
knowledge?

We believe that
we do have a useful and meaningful definition
of common knowledge.
However, a closer inspection of the situation is needed in order
to understand the subtle issues involved.
First of all, we shall see that only rather strong notions of
coordination in a distributed system require common knowledge.
Common knowledge corresponds to absolutely simultaneous coordination,
which is more than is necessary in many particular applications.
For many other types of coordination,
weaker states of knowledge suffice.
In the coming sections we investigate a variety of weaker states
of knowledge that are appropriate for many applications.
Furthermore, in many cases practical situations (and
practical distributed systems) can be faithfully modeled by a
simplified abstract model, in which common knowledge {\em is} attainable.
In such a case, when facts become common knowledge in the abstract
model it may be perfectly safe and reasonable to consider them to
be common knowledge when deciding on actions to be performed
in the actual system. We discuss this in greater detail in
Section~\ref{internal}.

\section{Common knowledge revisited}
 
\markj%
In Section~8 we showed that common knowledge is not attainable
in practical distributed systems under {\em any\/}
reasonable interpretation of
\markj %
knowledge (\ie in any epistemic interpretation).
Our purpose in the coming sections is to investigate what
states of knowledge {\em are\/} attainable in such systems.
For that purpose, we restrict our attention
to view-based interpretations of knowledge, since they seem
to be the
most appropriate for many applications in distributed systems.
Under view-based interpretations, it seems useful
to consider an alternative view of common knowledge.
 
Recall the children's state of knowledge of the fact {\bf m} in the muddy
children puzzle. If we assume that it is common knowledge that
all children comprehend~$\m$ simultaneously, then after the father
announces $\m$, the children attain $C\m$. However,
when they attain $C${\bf m} it is not the case that the children
learn
the infinitely many facts of the form
$E^k${\bf m} separately.
Rather, after the father speaks, the children are
in a state of knowledge~$S$
characterized by the fact that every child knows both
that $\m$ holds and
that $S$ holds.
Thus,~$S$ satisfies the equation
$$S\equiv E(\m \wedge S).$$
 
The fixed point axiom  of Section 6 says that under a
view-based interpretation, $C\subG \varphi$
is a solution for $X$ in an analogous fixed point equation, namely
$$X \equiv E\subG(\varphi \wedge X).$$
Now this equation has many solutions, including, for example,
both {\bf false} and $C\subG(\varphi\wedge \psi)$,
for any formula $\psi$.
$C\subG \varphi$ can be characterized
as being the {\em greatest fixed point\/}
of the equation; \ie a fixed point that is implied by all other
solutions.  (The {\em least fixed point\/}
 of this equation is
{\bf false}, since it
implies all other solutions.)
As our discussion of common knowledge in the case of the
muddy children puzzle suggests, expressing common knowledge
as a greatest fixed point of such an equation seems to correspond more
closely to the way it actually arises.
We sketch a semantics for a propositional view-based logic of
knowledge with fixed points in
 Appendix~A.
This alternative point of view, considering common knowledge
as the greatest fixed point of such an equation, will turn out
to be very useful when we attempt to define related variants of
common knowledge.
 
\section{$\eps$-common knowledge and $\D$-common knowledge}
\label{eps-ck section}
 
Since, strictly speaking,
common knowledge cannot be attained in practical
distributed systems, it is natural to ask what states of
knowledge {\em can\/} be obtained by the communication process.
In this section we consider what states of knowledge
are attained in systems in which communication delivery is guaranteed but
message delivery times are uncertain.
For ease of exposition, we restrict our attention to view-based
interpretations of knowledge here and in the next section.
 
We begin by considering {\em synchronous broadcast\/}
channels of communication;
\ie ones where every message sent is received by all processors,
and there are constants $L$ and $\eps$ such that all processors
receive the message between $L$ and  $L+\eps$ time units from the time
it is sent.  We call
$\eps$ the {\em broadcast spread\/} of such a channel.
Recall that the properties of the system
hold throughout all of its runs and hence
are common knowledge.
In particular, the properties of the broadcast channel are
common knowledge under any view-based interpretation.
 
Let us now consider the state of knowledge of the system
when a processor $p_i$
receives %
a broadcast message $m$.
Clearly $p_i$ knows that within an interval of $\eps$ time units around the
current time everyone (receives~$m$ and) knows $\snt(m)$.
But $p_i$ also knows that any other processor that
receives~$m$ will know that all processors will receive~$m$ within such
an $\eps$~interval.
Let us define {\em within an
$\eps$~interval, everyone knows~$\varphi$\/}, denoted
$\Ee\varphi$, to hold if there is an interval of~$\eps$ time units
containing the current time such that each processor comes to know~$\varphi$
at some point in this interval. Formally, we have:
$(\I,r,t)\sat\Ee\subG\varphi$ if there exists an interval~$I=[t',t'+\eps]$
such that $t\in I$ and for all $p_i\in G$ there exists~$t_i\in I$ for which
$(\I,r,t_i)\sat K_i\varphi$.
Let~$\psi$ be ``some processor has received~$m$''.
In a synchronous broadcast system as described above, we clearly have
that $\psi\imp\Ee\psi$ is valid.
 
We are thus in a state of knowledge that is analogous to common
knowledge;
here, however, rather than everyone knowing~$\varphi$ at the same instant,
they all come to know~$\varphi$ within an interval of~$\eps$ time units.
We call this the state of group knowledge
$\eps$-common knowledge, denoted~$\Ce$.  The formal definition of
$\Ce \subG \varphi$ is as the greatest fixed point
of the equation:
$$X\equiv \Ee\subG(\varphi \wedge X).$$
We refer the reader to Appendix~A for a rigorous definition.
The fact~$\psi$ above, stating that some processor received the message~$m$,
has the property that $\psi\imp \Ce\psi$. In addition, as $\psi\imp\snt(m)$
is valid, it is also the case that $\psi\imp\Ce\snt(m)$.
Thus, when some processor receives~$m$ it becomes $\eps$-common knowledge
that~$m$ has been sent.
 
As a straightforward consequence of its definition, $\Ce$ satisfies
the appropriate analogues of the fixed
point axiom~C1 and the induction rule~C2
of Section~6 (replacing~$E$ by~$\Ee$ and~$C$ by~$\Ce$).
Note that we did not define $\Ce \subG \varphi$ as an infinite
conjunction of $(\Ee\subG)^k\varphi$, $k \ge 1$.  While it is not
hard to show that $\Ce \subG \varphi$ implies this infinite conjunction,
it is not equivalent to it; however, giving a detailed counterexample
is beyond the scope of this paper.  (We give an example of a
similar phenomenon below.)  The fixed point definition is the
one that is appropriate for our applications.  Just as common
knowledge corresponds to simultaneous actions in a distributed
system, $\eps$-common knowledge corresponds to actions that are
guaranteed to be performed within $\eps$ time units of one
another.  This is what we get from the fixed point axiom~C1, which
does not hold in general for the infinite conjunction.
We are often interested in actions that are guaranteed to be performed
within a small time window.  For example,
in an ``early stopping'' protocol for
Byzantine agreement (\cf \cite{DRS}), all correct processors
are guaranteed to
decide on a common value within $\eps$ time units of each other.
It follows that once the first processor
decides, the decision value is $\eps$-common knowledge.%
\footnote{The situation there is in fact slightly
more complicated since only the correct processors are required to
decide; see \cite{MT} for definitions of knowledge appropriate
for such situations.}
 
There is one important special case where it can be shown that
the fixed point definition of $\Ce\subG \varphi$ is equivalent
to the infinite conjunction.  This arises when we restrict
attention to complete-history interpretations and
{\em stable\/} facts, facts that once true, remain true.
Many facts of interest in distributed systems applications,
such as ``$\phi$ held at some point in the past'', ``the
initial value of $x$ is 1'', or ``$\phi$ holds at time $t$
on $p_i$'s clock'', are stable.  If $\phi$ is stable, then
it is not hard to check that in complete-history interpretations,
we have that $\Ee\subG \phi$ holds iff $E\subG \phi$ will hold in $\eps$
time units.
As a straightforward consequence of this observation, we can show that
in complete-history interpretations, for a stable fact $\phi$ we do
have that $\Ce\subG \phi$ holds iff $(\Ee\subG) ^k \phi$ holds for
all $k \ge 1$.%
\footnote{We remark that in earlier versions of this paper, we
restricted attention to complete-history interpretations and
stable facts, and defined $\Ee\subG\phi$ as $\bigcirc^{\eps} E\subG\phi$,
where $\bigcirc^{\eps} \psi$ is true at a point $(r,t)$ iff $\psi$ is
true $\eps$ time units later, at $(r,t+\eps)$.  By the comments
above, our current definition is a generalization of our former
definition.}
 
It is not hard to verify that of the properties of S5, $\Ce$ satisfies
only A3 (positive introspection) and R1 (the rule of necessitation).
The failure of $\Ce$ to satisfy the knowledge axiom and the consequence
closure axiom can be traced to the failure of $\Ee$ to satisfy these
axioms.  The problem is that
$\Ee\varphi$ only requires that~$\varphi$ hold and be known at some (not
all!) of the points in the $\eps$~interval~$I$.
Indeed, it is not hard to construct an example in which
$\Ee\varphi\wedge\Ee\neg\varphi$ holds.
We remark that
if we restrict attention to stable facts and complete-history
interpretations, then consequence closure does hold for both $\Ee$
and $\Ce$

It is interesting to compare $\eps$-common knowledge with common knowledge.
Clearly, $C\varphi\imp\Ce\varphi$ is valid.
However, since synchronous broadcast channels are implementable in systems
where common knowledge is not attainable, the converse does not hold.
Thus, $\eps$-common knowledge is strictly weaker than common knowledge.
Moreover, note that while $C\varphi$ is a static state of knowledge,
which can be true of a point in time irrespective of its past or future,
$\Ce \varphi$ is a notion that is essentially temporal.
Whether or not it holds depends on what processors will know
in an~$\eps$ interval around the current time.

For any message $m$ broadcast on a channel with broadcast spread $\eps$,
the fact $\snt(m)$ becomes $\eps$-common knowledge
$L$ time units after $m$ is broadcast (in particular, as soon
as it is sent if $L=0$).  Upon receiving $m$, a processor $p_i$
knows that $\Ce \snt(m)$ holds, i.e.~$K_i \Ce \snt(m)$ holds.
Returning to R2 and D2's communication problem, we can view them as a
synchronous broadcast system, and indeed they attain
$\Ce \snt(m)$ immediately when R2 sends the message~$m$.
(Note that $L=0$ in this particular example; the interested reader
is invited to check that R2 and D2 in fact
achieve $\eps/2$-common knowledge of $\snt(m)$ at time $t_S+\eps/2$.)

Having discussed states of knowledge in synchronous broadcast
channels, we now turn our attention to systems in which
communication is {\em asynchronous}:
no bound on the delivery times of messages in the system exists.
Consider the state of knowledge of $\snt(m)$ in a system in which $m$ is
broadcast over an {\em asynchronous channel}:
a channel that guarantees that every message broadcast will
{\em eventually\/} reach every processor.
Upon receiving $m$, a processor knows $\snt(m)$,
and knows that every other processor either has already
received~$m$ or will
eventually receive~$m$.
This situation, where it is common knowledge
that if~$m$ is sent then
everyone will eventually know that~$m$ has been sent, gives rise to
a weak state of group knowledge which we call
{\em eventual common knowledge}.

We define {\em everyone in $G$ will eventually have
known~$\varphi$}, denoted~$\Ed\subG \varphi$,
to hold if for every processor in $G$
there is some time during the run at which
it knows~$\varphi$. Formally,
$(\I,r,t)\sat\Ed\subG\varphi$ if for all $p_i\in G$ there exists $t_i\ge 0$
such that $(\I,r,t_i)\sat K_i\varphi$.
We remark that if we restrict attention to stable facts $\phi$ and
complete-history interpretations, then $\Ed\subG \varphi$ is
equivalent to $\D E\subG \phi$, that is, eventually
everyone in $G$ knows $\phi$.%
\footnote{Formally, we take $\D \psi$ to be true at a point $(r,t)$
if $\psi$ is true at some point $(r,t')$ with $t' \ge t$.  In an
earlier version of this paper, we defined $\Ed\subG\phi$ as $\D E\subG
\phi$.  Again, by the comments above, our current definition is a
generalization of our former one.}
We define
the state of $\D$-{\em common knowledge\/} (read {\em eventual\/}
common knowledge),
denoted by $\Cd$, by taking $\Cd\subG \varphi$ to be the
 greatest fixed point of the equation:
$$X \equiv\ \Ed\subG(\varphi\wedge X).$$
Notice that we
again used the fixed point definition rather than one in
terms of infinite conjunction of $(\Ed\subG)^k \phi$, $k \ge1$.
Our definition implies the infinite conjunction but,
as we show by example below, it is not equivalent to the
infinite conjunction, even if we restrict to stable facts
and complete-history interpretations.
 
Our motivation for considering the fixed point definition
is the same as it was in the case of $\eps$-common knowledge.
The fixed point definition gives us analogues to C1 and C2;
as a consequence,
$\D$-common knowledge corresponds to events that
are guaranteed to take place at all sites {\em eventually}.
For example, in some of the work on
variants of the Byzantine Agreement problem discussed in the
literature (\cf \cite{DRS}), the kind of agreement sought is
one in which whenever a correct processor decides
on a given value, each other correct processor is guaranteed to
{\em eventually\/} decide on the same value.
The state of knowledge of the decision value that the
processors attain in such circumstances is
$\D$-common knowledge.
Also, in asynchronous error-free broadcast channels, a processor
knows that $\snt(m)$ is $\D$-common knowledge when it receives the
message $m$.

$\Cd\subG$ is the weakest temporal notion of common knowledge that we have
introduced. In fact, we now have a hierarchy of the temporal notions of
common knowledge.  For any fact~$\varphi$ and
$\eps_1 \le \cdots \le \eps_k \le \eps_{k+1} \le \cdots\,$, we have:
$$C\subG\varphi\imp\C^{\eps_1}\subG\varphi\imp\cdots\imp
C^{\eps_k}\subG\varphi\imp
C^{\eps_{k+1}}\subG\varphi\imp\cdots\imp\Cd\subG \varphi.$$

We next consider how $\Ce$ and $\Cd$
are affected by communication not
being guaranteed.
\markj%
Recall that  Theorem~\ref{Lemma3} %
implies that if communication
is not guaranteed, then common knowledge is independent of
the communication process.  A fact only becomes common knowledge
if it becomes common knowledge in the absence of messages.
Interestingly, the obvious analogue of Theorem~\ref{Lemma3}
does not hold for $\Ce$ and $\Cd$.
Indeed, it is possible to construct a situation in which $\Ce\varphi$
is attained {\em only\/} if communication is not
sufficiently successful.
For example, consider a system consisting of R2 and D2 connected by
a two-way link. Communication along the link is not guaranteed,
R2 and D2's clocks are perfectly synchronized, and
\markj%
both of them run the following protocol: {\sl At time 0, send
the message ``OK''. For all natural numbers $k>0$,
if you have received~$k$ ``OK''
messages by time $k$ on your clock, send an ``OK'' message at time $k$;
otherwise, send nothing.}
Let $\psi=$``it is time $k$ where $k \ge 1$ and some message sent at or
before
time $k-1$  was not delivered within one time unit.''
\markj%
Assume a complete-history interpretation for this system and fix
$\eps = 1$.
It is easy to see that
$\psi\imp\Ee\psi$
is valid in this system.  For suppose that
at time $k$ the fact $\psi$ holds because one of R2's messages was not
delivered to D2.  D2 knows $\psi$ at time $k$ and,
according to the protocol,
will not send a message to R2 at time $k$.  Thus, by time $k+1$,
R2 will also know $\psi$ (if it didn't know it earlier).
The induction rule implies that $\psi\imp\Ce \psi$ is also valid
in the system.
If $r$ is a run of the system where no messages are received,
then it is easy to see that $\psi$ holds at $(r,1)$, and
hence so does $\Ce \psi$.  However, $\Ce \psi$ does not hold at
$(r',1)$ if $r'$ is a run where all messages are delivered
within one time unit.
(The same example works for $\Cd \psi$.)

In the example above, successful communication
in a system where communication is not guaranteed can prevent
$\Ce\subG \psi$ (resp.\ $\Cd \subG\psi$) from holding.
However, the following theorem shows that
we can get a partial analogue to Theorem~\ref{Lemma3}
for $\Ce$ and $\Cd$.
Intuitively, it states that if $\Ce\subG\psi$ (resp.\ $\Cd\subG\psi$)
does not hold in the absence of successful communication,
then $\Ce\subG\psi$ (resp.\ $\Cd\subG\psi$) does not hold regardless of
how successful communication may turn out to be.
More formally,
 
\begin{theorem}
\label{Theorem8}
Let $R$ and~$G$ be as in Theorem~\ref{Lemma3}, and let $\I$ be a
view-based interpretation.
Let $r^-$ be a run of $R$ where no messages are received.  If
$(\I,r^-,t) \not\sat \Ce\subG \varphi$
(resp.~$(\I,r^-,t) \not\sat \Cd\subG\varphi$) for all times $t$, then
$(\I,r,t)\not\sat\Ce\subG
\varphi$ (resp.\ $(\I,r,t)\not\sat\Cd \subG \varphi$) for all runs $r$
with the same initial configuration and the same clock readings
as $r^-$ and all
times $t$.
\end{theorem}
 
\begin{proof}
\markj%
We sketch the proof for $\Ce\subG \varphi$; the proof for $\Cd\subG \varphi$
is analogous.
We assume that all runs mentioned in this proof have
the same initial configuration and the same clock readings as $r^-$.
If $r$ is a run such that $\Ce\subG \varphi$ holds at some
point in $r$, let $t_j(r)$ be the first time in~$r$ that processor~$
p_j\in G$
knows~$\Ce\subG\varphi$. Let $\hat t (r) =\max\{t_j (r) :p_j\in G\}$,
and let $d(r)$ be the number of messages that are
received in $r$ up to (but not including) $\hat t (r)$.
We show by induction on $k$ that if $r$ is a run such that $\Ce\subG
\varphi$ holds at some point in $r$, then
$d(r) \ne k$.   This will show that in fact $\Ce\subG\varphi$ can never
hold.
 
If $d(r) = 0$ and $\Ce\subG \varphi$ holds at some point in $r$,
choose some
$p_i\in G$ and let $t_i = t_i(r)$.
Then we have that $(\I,r,t_i) \sat K_i\Ce\subG \varphi$.
Clearly $h(p_i,r,t_i) = h(p_i,r^-,t_i)$, so $(\I,r^-,t_i) \sat K_i
\Ce\subG \varphi$.  By the knowledge axiom, we have that $(\I,r^-,t_i) \sat
\Ce\subG \varphi$, contradicting the hypothesis of the theorem.
 
For the inductive step,
assume that $d(r) = k+1$ and
let $\hat t=\hat t(r)$.
We now proceed as in the proof of Theorem~\ref{Lemma3}.
Let~$p_j$ be a processor receiving the last
message received in $r$ before time~$\hat t$.
Let~$t'$ be the time at which $p_j$ receives this message.
Let $p_i$ be a processor in $G$ such that $p_i \ne p_j$
and let $t_i = t_i(r)$.
Since communication is not guaranteed, there exists a run
$r'$ extending $(r,t')$ such that (1) no messages are received in $r'$
at or after time $\hat t$, (2) $h(p_i,r,t'') = h(p_i,r',t'')$
for all $t'' \le \hat t$, and (3)
all processors $p_k\ne p_i$ receive no messages in the interval
$[t', \hat t)$.  By construction, at most $k$ messages are received
altogether in $r'$, so $d(r') \le k$.
By the induction hypothesis we have that
$(\I,r',t'')\sat \neg \Ce\subG\varphi$ for all $t''$.  It follows that
$(\I,r',t_i) \sat \neg K_i \Ce \subG\varphi$.  But since we assumed
$(\I,r, t_i) \sat K_i \Ce \subG \varphi$ and $h(p_i,r, t_i) = h(p_i,r',t_i)$,
this gives us a contradiction.
\qed
\end{proof}
 
We can now use Theorem~\ref{Theorem8} to prove an analogue to
Corollary~\ref{Proposition5.1},
which shows that if communication is not guaranteed, then
there is no protocol for eventually coordinated attack.
 
\begin{proposition}
In the coordinated attack  problem,
any protocol that guarantees that whenever either party attacks
the other party will {\em eventually\/} attack,
is a protocol in which necessarily neither party attacks.
\end{proposition}
 
\begin{proof}
The proof is analogous to that of
Corollary~\ref{Proposition5.1}. Assume that $(P_A, P_B)$ is a joint
protocol that guarantees
that if either party attacks then they both eventually attack, and let
$R$ be the corresponding system.
\markj%
Let $\psi=$``At least one of the generals has started attacking''.
We first show that when either general attacks, then eventual
 common knowledge of~$\psi$ must hold.
\markj%
Since the protocol guarantees that whenever one general attacks the
other one eventually attacks, it is easy to see that a
general that has decided to attack
knows~$\psi$ and knows that eventually both generals will
know~$\psi$. Thus, by the induction rule for~$\Cd$, when a general
attacks $\Cd\psi$ holds.
 Since in every run of the protocol in which no messages
are received no party attacks (and hence neither $\psi$ nor $\Cd \psi$
hold in such runs),
by Theorem~\ref{Theorem8}, the protocol $(P_A, P_B)$ guarantees that
neither general will ever attack.
\qed
\end{proof}

Theorem~\ref{Theorem8} allows us to construct an example in which
the infinite conjunction of~$(\Ed)^k\varphi$ holds, but $\Cd\varphi$
does not. In the setting of the coordinated attack problem,
Let~$\varphi$ be ``{\sl General~A is in favor of attacking}''.
Consider a run in which all messengers arrive safely, and
messages are acknowledged ad infinitum. Clearly, assuming a
complete-history interpretation, for all~$k$
it is the case that~$E^k\varphi$ holds after the~$k$th message is
delivered. It follows that~$(\Ed)^k\varphi$ holds at time~0.
However, Theorem~\ref{Theorem8} implies that~$\Cd\varphi$
never holds in this run.
It follows that~$\Cd\varphi$ is not equivalent to the infinite conjunction
of~$(\Ed)^k\varphi$ even in the case of stable facts~$\varphi$ and
complete-history interpretations.
 
Recall that the proof that unreliable communication cannot affect
what facts are common knowledge carried over to
(reliable) asynchronous communication.
Our proof in Theorem~\ref{Theorem8} clearly does not carry over. In fact,
a message broadcast over a reliable asynchronous channel
{\em does\/} become eventual common knowledge. However, it is possible
to show that asynchronous
channels cannot be used in order to attain $\eps$-common knowledge:
 
\begin{theorem}
\markj%
Let $R$ be a system with unbounded delivery times and let
$\vert G \vert \ge 2$. %
Suppose there is
some run $r^-$ in $R$ in which no message are delivered in the
interval $[0,t+\eps)$ such that  $(\I,r^-,t) \not\sat \Ce\subG\psi$.
Then for all runs $r$ in $R$ with the same initial configuration and
the same clock readings as $r^-$, we have
$(\I,r,t)\not\sat\Ce \subG \psi$.
\end{theorem}
 
\noindent{\bf Sketch of Proof:}\quad The proof essentially follows the
proof of Theorems~\ref{Lemma3} and~\ref{Theorem8}.  We proceed by
induction on $d(r)$, the number of messages received in $r$ up to
time $t$.  Details are left to the reader.  \qed
 
Thus, asynchronous communication channels are of no use for
coordinating actions that are guaranteed
to be performed at all sites within a predetermined fixed time bound.

\section{Timestamping: using relativistic time}
 
{\em Real\/} time is not always the appropriate notion of time to
consider in a distributed system.
Processors in a distributed system often do not have access
to a common source of real time, and their clocks do not show identical
readings at any given real time.
Furthermore, the actions taken by the processors rarely actually
depend on real time.
Rather, time is often used mainly
for correctly sequencing events at the different sites
and for maintaining  ``consistent''  views
of the state of the system.
In this section we consider states of knowledge relative
to {\em relativistic\/} notions of time.
 
Consider the following scenario:
R2 knows that R2 and D2's clock differ by at most~$\delta$, and that any
message R2 sends D2 will arrive within $\eps$ time units.
R2 sends D2 the following message $m'$:
 
\begin{quote}
``This message is being sent at $t_S$ on R2's clock,
and will reach D2 by $t_S+\eps+\delta$ on both clocks; $m$.''
\end{quote}
 
Let us denote $t_S+\eps+\delta$ by $T_0$.
Now, at time $T_0$ on his clock, R2 would like to claim that
$\snt(m')$ is common knowledge. Is it? Well, we know by now
that it is not,
but it is interesting to analyze this situation.
Before we do so, let us introduce a relativistic formalism
for knowledge, which we call {\em timestamped\/} knowledge:
We denote ``at time $T$ on his clock, $p_i$ knows~$\varphi$''
by $K_i^\ts \varphi$.
$T$ is said to be the {\em timestamp\/} associated with this knowledge.
We then define
$$E^\ts\subG \varphi\equiv \bigwedge_{p_i\in G} K_i^\ts \varphi.$$
$E^\ts \varphi$ corresponds to
everyone knowing~$\varphi$ individually at time $T$ on their own clocks.
Notice that for $T_0$ as above, $\snt(m')\imp\Etz \snt(m')$.
It is natural to define the corresponding relativistic
variant of common knowledge, $\C^\ts$,
which we call {\em timestamped\/} common knowledge, so that
$\C^\ts\subG \varphi$ is the greatest fixed point of the equation
$$X \equiv  \Ets\subG(\varphi \wedge X).$$
So, in any run where the message $m'$ is sent,
R2 and D2 have timestamped common knowledge of $\snt(m')$ with
timestamp $T_0$.
It is easy to check that $\Cts$ satisfies the fixed point axiom and
the induction rule, as well as all of the axioms of~S5 except for the
knowledge axiom.  In this respect, $\Cts$ resembles $C$ more closely
than~$\Ce$ and~$\Cd$ do.
 
It is interesting to investigate how the
relativistic notion of timestamped
common knowledge relates to the notions of common knowledge,
$\eps$-common knowledge, and $\D$-common knowledge.
Not surprisingly, the relative behavior of the clocks in the
system plays a crucial role in determining the meaning of~$\Cts$.
 
\begin{theorem}
\label{Theorem11}
For any fact~$\varphi$ and view-based interpretation,
\begin{itemize}
\item[(a)] if it is guaranteed that all clocks show identical
times, then at time~$T$ on any processor's clock,
$\Cts\subG \varphi\equiv C\subG\varphi$.
\item[(b)]
if it is guaranteed that all clocks are within $\eps$ time units of
each other, then at time~$T$
on any processor's clock, $\Cts\subG \varphi\imp\Ce\subG \varphi$.
\item[(c)] if it is guaranteed that
each local clock reads~$T$ at some time,
then   %
$\Cts\subG \varphi\imp \Cd\subG \varphi$.
\qed
\end{itemize}
\end{theorem}

Theorem~\ref{Theorem11}
gives conditions under which
$\Cts$ can be replaced by $C$, $\Ce$, and $\Cd$.
A weak converse of Theorem~\ref{Theorem11} holds as well.
Suppose the
processors are able
to set their clocks to a commonly agreed upon time~$T$ when
they come to know $C\subG\varphi$ (resp.\ come to know
$\Ce \subG \varphi$, $\Cd\subG \varphi$).
Then it is easy to see that whenever $C\subG\varphi$
(resp.\ $\Ce \subG \varphi$, $\Cd\subG \varphi$) is attainable,
so is $\Cts \subG \varphi$.
 
In many distributed systems timestamped common knowledge seems to be
a more appropriate notion to reason about than ``true'' common knowledge.
Although common knowledge cannot be attained in practical systems,
timestamped common knowledge is attainable in many cases of interest
and seems to correspond closely to the relevant phenomena
with which protocol designers are confronted.
For example, in distributed protocols that work in phases,
we speak of the state of the system at the beginning of
phase~2, at the end of phase~$k$, and so on.
It is natural to think of the phase number as a ``clock'' reading,
and consider knowledge about what holds at the different phases as
``timestamped'' knowledge, with the phase number being the timestamp.
In certain protocols for Byzantine agreement, for example,
the nonfaulty processors attain common knowledge of the decision
value at the end of phase~$k$ (\cf \cite{DM,MT}).
In practical systems in which the phases do not end
simultaneously at the different sites of the system, the processors
can be thought of as actually attaining timestamped common
knowledge of the decision value,
with the timestamp being ``the end of phase~$k$''.
  Indeed, protocols like the atomic broadcast protocol
of \cite{CASD} are designed exactly for the purpose of attaining
timestamped common knowledge.  (See \cite{NT} for more discussion
of timestamped common knowledge.)

\section{Internal knowledge consistency}\label{internal}
 
We have seen that common knowledge closely corresponds to the ability
to perform {\em simultaneous\/} actions.
In the last few sections we introduced a
number of related states of knowledge corresponding to weaker
forms of coordinated actions.
Such weaker forms of coordination are often sufficient for many practical
applications. This helps explain the paradox of the happy existence of
practical distributed systems despite the apparent need for
\markj%
common knowledge and the negative results of Theorem~\ref{PracProp}.
 
However, there are situations where we act as if -- or we would like
to carry out our analysis as if -- we had true common knowledge, not a
weaker variant. For example, in the muddy children puzzle, even though
simultaneity may not be attainable, we want to assume that
the children do indeed have common knowledge of the father's statement.
As another example, consider a protocol that proceeds in phases,
in which it is guaranteed that no processor will ever
receive a message out of phase.
In many cases,
all the aspects of this protocol that we may be interested in
are faithfully represented if we model the system as if it were truly
synchronous: all processors switch from one phase to the next
simultaneously.

Intuitively, in both cases, the assumption of common knowledge seems
to be a {\em safe\/} one, even if it is not quite true.  We would like to
make this intuition precise.  Recall that an epistemic interpretation
is one that specifies what a processor believes at any given point as a
function of the processor's history at that point.  An epistemic
interpretation~$\I$ is a knowledge interpretation if it is {\em knowledge
consistent}, \ie if it has the property that
whenever $(\I,r,t) \sat K_i \varphi$ then also
$(\I,r,t) \sat \varphi$.
Now an epistemic interpretation that is not knowledge consistent may
nevertheless be {\em internally knowledge consistent},
which intuitively means that the processors
never obtain information from within the system that would
contradict the assumption that the epistemic interpretation
is in fact a knowledge interpretation.
In other words, no processor ever has information that implies that
the knowledge axiom $K_i\varphi\imp\varphi$ is violated.
More formally, an epistemic interpretation~$\I$ for a system~$R$ is said
to be {\em internally knowledge consistent\/} if there is a subsystem
$R'\subseteq R$ such that $\I$ is a knowledge interpretation  when
restricted to $R'$,
and for all processors~$p_i$ and points $(r,t)$ of~$R$, there is
a point $(r',t')$ in~$R'$ such that $h(p_i,r,t)=h(p_i,r',t')$.
 
Given that epistemic interpretations ascribe knowledge (or, perhaps
more appropriately in this case, beliefs) to processors as
a function of the processors' histories, the above definition implies
that
whenever a processor is ascribed knowledge of a certain fact at a point
of~$R$, then as far as any events involving this processor at the
current and at any future time are concerned, it is consistent to assume
that the fact does indeed hold.

Using the notion of internal knowledge consistency, we can make
our previous intuitions precise.
When analyzing the muddy children puzzle, we
assume that the children will never discover that they did not hear and
comprehend the father's statement simultaneously.  We take the set~$R'$
from the definition of internal knowledge consistency here to be
precisely the set of runs where they did hear and comprehend the
father's statement simultaneously.  Similarly, in the case of the
protocol discussed above, the set $R'$ is the set where all
processors advance from one phase to the next truly simultaneously.
It now also makes sense to say that under reasonable conditions
processors can safely use an ``eager'' protocol corresponding to the
eager epistemic interpretation of Section 8,
in which processors act as if they had common knowledge, even
though common knowledge does not hold.
It is possible to give a number of conditions on the ordering of events
in the system that will ensure that it will be internally knowledge
consistent for the processors to act as if they have common knowledge.

For further discussion on internal knowledge consistency, see
the recent paper by Neiger \cite{Nei}.
 
\section{Conclusions}\label{conc}

In this paper, we have tried to bring out the important role of
reasoning about knowledge in distributed systems.
We have shown that reasoning about the knowledge of a group
and its evolution
can reveal subtleties that may not
otherwise be apparent, can sharpen our understanding of basic issues,
and can improve the high-level reasoning required in the design and
analysis of distributed protocols and plans.
 
We introduced a number of states of group knowledge, but focused
much of our attention on one particular state, that of common knowledge.
We showed that, in a precise sense, common knowledge is a prerequisite
for agreement.  However, we also showed that in many practical systems
common knowledge is not attainable.
This led us to consider three
variants of common knowledge --- $\eps$-common knowledge, eventual
common knowledge, and timestamped common knowledge ---
that are attainable in practice, and
may suffice for carrying out a number of actions.
The methodology we introduce for constructing these variants of common
knowledge, involving the fixed-point operator, can be used to construct
other useful variants of common knowledge.  Indeed, recent papers
have introduced {\em concurrent common knowledge\/} \cite{PT},
{\em probabilistic common knowledge\/} \cite{FH3}, and
{\em polynomial time common knowledge} \cite{Mos},
 using this methodology.
 
There is clearly much more work to be done in terms of gaining
a better understanding of knowledge in distributed systems.
This paper considers a general model of a distributed system.
It would also be useful to consider
knowledge in distributed systems with particular properties.
The work of Chandy and Misra \cite{ChM} is an interesting study of this
kind (see \cite{DM,FHV2,Had} for other examples).
We  carried out a knowledge-based analysis of the coordinated
attack problem here.
Since this paper first appeared, a number of other problems, including
Byzantine agreement, distributed commitment, and mutual exclusion,
have been analyzed in terms of knowledge
(see \cite{ChM,DM,Had,HZ,Maz,MT,NT}).
Such knowledge-based analyses both shed light on the problem being
studied and improve our understanding of the methodology.
More studies of this
kind would further deepen our understanding of the issues involved.

Another general direction of research is that of using knowledge
for the specification and verification of distributed systems.
(See \cite{KT} for an initial step in this direction.)
Formalisms based on knowledge may prove to be a powerful tool
for specifying and verifying protocols, and may also be readily
applicable to the synthesis of protocols and plans.
Temporal logic has already proved somewhat successful
in this regard \cite{EC,MW}.

Our analysis of the muddy children puzzle and the coordinated attack
problem, as well as the work in \cite{MDH,HF,DM,MT} illustrate
how subtle the relationship between knowledge, action, and communication
in a distributed system can be.
In this context, Halpern and Fagin (\cf \cite{HF}) look at
{\em knowledge-based protocols},
which are protocols in which a processor's actions are explicitly
based on the processor's knowledge. This provides an interesting
generalization of the more standard notions of protocols.
 
In the long run, we hope that a theory of
knowledge, communication, and action
will prove rich enough to provide general foundations
for a unified theoretical treatment of distributed systems.
Such a theory also promises to shed light on
aspects of knowledge that are relevant to related fields.
 
\bigskip \gb\noindent{\bf  Acknowledgements:}\quad
This work evolved from work the authors did with Danny Dolev on \cite{MDH}.
Many people commented on different versions of this work.
Of special value
were comments by Dave Chelberg, Steve Deering, Cynthia Dwork,
Ron Fagin, Vassos Hadzilacos, Danny Lehmann, Yoni Malachi, Tim Mann,
Andres Modet, Gil Neiger, Jan Pachl, Derek Proudian, Stan Rosenschein,
 Yoav Shoham,
Ray Strong, Moshe Vardi, Joe Weening, and Lenore Zuck.  Jan Pachl
suggested the term ``distributed knowledge'', to replace the term
``implicit knowledge'' that we had been using.
We would particularly like
to thank Gil Neiger and Lenore Zuck for an outstanding  job of
refereeing, well
beyond the call of duty.

\section*{Appendix A}
 
In this appendix we present a logic with a greatest fixed point
operator and illustrate how  common knowledge and variants
of common knowledge can be formally defined as greatest fixed points.
Our presentation  follows that of Kozen \cite{Koz1}.

Intuitively, given a system~$R$, a formula~$\psi$
partitions the points of~$R$ into two sets: those that satisfy~$\psi$,
and those that do not. We can identify a formula with the set
of points that satisfy it. In order to be able to define
fixed points of certain formulas, which is our objective in this appendix,
we consider formulas that may contain a free variable whose values
range over subsets of the points of~$R$.
Once we assign a set of points to  the free variable, the formula
can be associated with a set of points
in a straightforward way (as will be shown below).
Thus, such a formula can be viewed as
a function from subsets of~$R$ to subsets of~$R$.
(A formula with no free variable is then considered a constant function,
yielding the same subset regardless of the assignment.)
 
Before we define the logic more formally, we need to review a number
of relevant facts about fixed points.
Suppose $S$ is a set and $f$ is a function mapping
subsets of~$S$ to subsets of~$S$. A subset~$A$ of~$S$ is said to be
a {\em fixed point\/} of~$f$ if $f(A)=A$. A {\em greatest\/}
(respectively, {\em least\/}) fixed point of $f$ is a set $B$ such that
$f(B) = B$, and if $f(A) = A$, then
$A\subseteq B$ (resp.\ $B\subseteq A$). It follows that if $f$ has
a greatest fixed point~$B$, then $B=\bigcup\{A\,:\,f(A)=A\}$.
The function~$f$ is said to be {\em monotone increasing\/} if
$f(A)\subseteq
f(B)$ whenever $A\subseteq B$ and {\em monotone decreasing\/} if $f(A)
\supseteq f(B)$ whenever \mbox{$A\subseteq B$}. The Knaster-Tarski theorem
(\cf \cite{Tar}) implies
that a monotone increasing function has a greatest (and a least) fixed
point. Given a function $f$ and a subset $A$, define $f^0(A)=A$ and
$f^{i+1}(A)=f(f^i(A))$.
$f$ is said to be {\em downward continuous\/} if
$f(\bigcap_iA_i)=\bigcap_if(A_i)$ for all
sequences $A_1,A_2,\ldots$ with $A_1 \supseteq A_2 \supseteq \ldots$.
Given a monotone increasing and downward continuous
function $f$
it is not hard to show that
the greatest fixed point of~$f$ is the set $\bigcap_{k<\omega}f^k(S).$
We remark that if $f$ is monotone increasing but not downward
continuous, then we can still obtain the greatest fixed points
of $f$ in this fashion, but we have to extend the construction by
defining $f^{\alpha}$ for all ordinals $\alpha$.%
\footnote{We can similarly define a
function~$f$ to be {\em upward continuous\/} if
$f(\bigcup_iA_i)=\bigcup_if(A_i)$ for all sequences $A_1, A_2, \ldots$
with $A_1 \subseteq A_2 \subseteq \ldots$.
For monotone increasing
upward continuous functions~$f$, the least fixed point of $f$ is
$\bigcup_{k<\omega}f^k(\emptyset)$.  Again, to get least fixed points
in the general case, we have to extend this construction through
the ordinals.}
 
We are now in a position to formally define our logic.
We start with a set
$\Phi=\{P,Q,P_1,\ldots\}$ of primitive propositions and
a single propositional variable~$X$.
We form more complicated formulas by
allowing the special formula $true$ and then
closing off under
conjunction, negation, the modal operators $K_i$, $E\subG$,
$\Ee\subG$, and $\Ed\subG$ for every group~$G$
of processors, and the greatest fixed point operator $\nu X$.
Thus, if $\varphi$ and $\psi$ are formulas, then so are
$\neg \varphi$, $\varphi \land \psi$, $K_i\varphi$, $E\subG\varphi$,
$\Ee\subG\varphi$, $\Ed\subG\varphi$, and $\nu X. \varphi$
(read ``the greatest fixed point of $\varphi$ with respect to $X$'').
However, we place a syntactic restriction, described below,
on formulas of the form $\nu X. \varphi$.
 
Just as $\forall x$ in first-order logic binds occurrences of $x$,
$\nu X$ binds
occurrences of
$X$.
Thus, in a formula such as $X \land \neg \Ee\subG ( \nu X.[ X \land
(K_1 X \land K_2 X)])$, the first occurrence of $X$ is free, while
the rest are bound.  We say that a free occurrence of $X$ in
a formula $\varphi$ is {\em positive\/} if it is in the scope of an even
number of negation signs, and {\em negative\/} if it is in the scope of
an odd number of negation signs.  Thus, in a formula such as $X \land
\neg
K_1 X$, the first occurrence of $X$ is positive while the second is
negative.  The restriction on formulas of the form $\nu X. \varphi$ is
that all free occurrences of $X$ in $\varphi$ must be positive;
the point of this restriction will be explained below.
 
The next step is to associate with each formula a function.
Given a distributed system represented by its set of runs $R$,
let $S=R\times [0,\infty)$.
A model ${\cal M}$ is a triple $(S,\pi,v)$, where $S$ is as above,
$\pi$ associates a truth assignment to the primitive propositions with
each point in $S$,
and $v\,:\,\{1,\ldots,m\}\times S\rightarrow \Sigma$
is an assignment of views (from a set of states $\Sigma$) to
the processors at the points of $S$.
We now associate
with each formula
$\varphi$ a function $\varphi^\cM $ from subsets of $S$ to subsets of $S$.
Intuitively, if no occurrences of $X$ are free in
$\varphi$, then $\varphi^\cM$ will be a constant function, and
$\varphi^\cM(A)$ will be the set of points where $\varphi$ is true
(no matter how we choose $A$).  If $X$ is free in $\varphi$,
then $\varphi^\cM(A)$ is the set of points where $\varphi$ is true if
$A$ is the set of points where $X$ is true.  We define
$\varphi^\cM(A)$ by induction on the structure of $\varphi$ as follows:
 
\begin{itemize}
\item[(a)] $X^\cM (A)=A$ (so $X^\cM$ is the identity function).
\item[(b)] $P^\cM (A)=\{s\in S\,:\, \pi(s)(P)={\bf true}\}$ for a
primitive proposition $P$
\item[(c)] $true^\cM (A) = S$.
\item[(d)] $(\neg \varphi)^\cM (A)=S- \varphi^\cM (A)$.
\item[(e)] $(\varphi\wedge \psi)^\cM (A)=
                                \varphi^\cM (A)\cap \psi^\cM (A)$.
\item[(f)] $(K_i\varphi)^\cM (A)=\{(r,t)\in S:\,
\hbox{for all $(r',t')\in S$,
$v(p_i,r,t)=v(p_i,r',t')$ implies $(r',t')\in \varphi^\cM (A)$}\}$.

\item[(g)] $(E\subG\varphi)^\cM (A)=
\cap_{i \in G}(K_i\varphi)^\cM(A)$.
\item[(h)] $(\Ee\subG\varphi)^\cM (A)=\{(r,t)\in S:\,
\mbox{there exists an interval }
I=[t',t'+\eps] {\rm\ with \ } t\in I,\ {\rm such \ that\ }\\
\forall p_i\in G\ \exists t_i\in I\ (
(r,t_i)\in (K_i\varphi)^\cM(A))\}$.
\item[(i)] $(\Ed\subG\varphi)^\cM (A)=\{(r,t)\in S:\,
\forall p_i\in G\ \exists t_i\ (
(r,t_i)\in (K_i\varphi)^\cM(A))\}$.
\item[(j)] $(\nu X.\varphi)^\cM (A)=
                               \bigcup\{B:\,\varphi^\cM (B)=B\}$.
\end{itemize}
 
Now by an easy induction on the structure
of formulas we can prove the following facts:
\begin{enumerate}
\item
If $\varphi$ is a formula in which all free occurrences of $X$
are positive (resp. negative), then $\varphi^\cM$ is monotone
increasing (resp. monotone decreasing).  Note that
our syntactic restriction on formulas of the form~$\nu X.\varphi$
guarantees that for a well-formed formula of this form,
the function $\varphi^\cM$ is monotone increasing.
As a consequence, $(\nu X.\varphi)^\cM(A)$
is the greatest fixed point of the function $\varphi^\cM$.
\item
If $\varphi$ is a formula with no free variables, then $\varphi^\cM$ is
a constant function.  In particular,
observe that $(\nu X. \varphi)^\cM$ is necessarily a constant function
(the definition shows that $(\nu X.\varphi)^\cM (A)$ is independent
of the choice of $A$).  As well, it is easy to check that
if $\varphi$ is a valid formula
such as $\neg(P \land \neg P)$, then $\varphi^\cM(A) =S$.
\item
For formulas in which the variable $X$ does not appear (so, in
particular, for formulas not involving the greatest fixed point
operator), $\varphi^\cM(A) = \{(r,t) : (\I_v,r,t)\sat \varphi \}$,
where
$\I_v$ is the view-based interpretation associated with the view function~$v$.
(Again, this is true for any choice of $A$, since by the
previous observation, $\varphi^\cM$ is a constant function if there is
no occurrence of $X$ in $\varphi$.)
Thus, if we define $(\cM,r,t) \sat \varphi$ iff $(r,t) \in
\varphi^\cM(\emptyset)$, then this definition
extends
our previous definition (in that for formulas in which the variable $X$
does not appear, we have $(\cM,r,t) \sat \varphi$ iff
$(\I_v,r,t) \sat \varphi$).
\end{enumerate}
 
Given the machinery at our disposal, we can now formally define
$C\subG\varphi$ as $\nu X.E\subG(\varphi\wedge X)$, define
$C^\epsilon\subG \varphi$ as $\nu X.\Ee\subG(\varphi \wedge X)$, and
define $\Cd\subG \varphi$ as $\nu X.\Ed\subG(\varphi\wedge  X)$.
It follows from our characterization of greatest fixed points
of downward continuous functions that if $\varphi^\cM$ is downward
continuous, then $\nu X. \varphi$ is
equivalent to $\varphi_0 \wedge \varphi_1 \wedge \ldots$, where
$\varphi_0$ is ${true}$, $\varphi_{i+1}$ is
$\varphi[\varphi_i/X]$, and $\varphi[\psi/X]$ denotes the result of
substituting~$\psi$ for the free occurrences of $X$ in $\varphi$.
It is easy
to check that $(E\subG(\varphi\wedge X))^\cM$ is downward continuous if
$\varphi^\cM$ is downward continuous.  In particular, if $\varphi$ has
no free occurrences of $X$ (so that $\varphi^\cM$ is constant), it
follows that we have:
$$C\subG\varphi\equiv
E\subG\varphi \wedge E\subG(\varphi \wedge E\subG\varphi)
\wedge E\subG(\varphi
\wedge E\subG(\varphi \wedge E\subG\varphi))
\wedge \ldots.%
\footnote{Note that
the formula on the right-hand side of the equivalence is not
in our language, since we have not allowed infinite conjunctions.
However, we can easily extend the language to allow infinite
conjunctions in the obvious way so that the equivalence holds.}
$$
Since $E\subG(\psi_1 \wedge \psi_2)
\equiv (E\subG\psi_1 \wedge E\subG\psi_2)$
it follows that
$$C\subG\varphi\equiv E\subG\varphi\wedge E\subG
E\subG\varphi\wedge\cdots.$$

However, $(\Ee\subG(\varphi\wedge X))^\cM$ and
$(\Ed\subG(\varphi\wedge X))^\cM$ are {\em not\/} necessarily
downward continuous. The reason that $(\Ed\subG(\varphi\wedge X))^\cM$
is not  downwards continuous is that an infinite collection of facts
can each {\em eventually\/} hold,
without them necessarily all holding simultaneously at some point.
We have already seen one example of this phenomenon
 in Section~\ref{eps-ck section}.  For another example, suppose we are
working in a system with an unbounded global clock, and let
$A_i = (current\_time>i)^{\cM}$.  Since the clock is unbounded, it follows   that
$A_i\ne\emptyset$ for
all $i$, but
$\cap_i A_i=\emptyset$.
Taking $\psi$ to be the formula
$\Ed(\varphi\wedge X)$,
it is easy to see that $(r,0)\in \psi^{\cM}(A_i)$ for all $i$, and hence
$\cap_i(\psi^{\cM}(A_i)) \ne \psi^{\cM}(\cap_i A_i)$.
 
We can construct a similar example
in the case of~$\Ee$, because we have taken time to
range over the reals.  For example, if we take $x_i$ to be an infinite
sequence of real numbers coverging from below to $\epsilon$, take
$A_i = (current\_time \in
(x_i,\epsilon))^{\cM}$, and now take $\psi$ to
be the formula $\Ee(\varphi\wedge X)$, then again we have
$\cap_i A_i = \emptyset$, and $(r,0) \in \cap_i
(\psi^{\cM}(A_i))$.
This example does depend crucially on the fact that time ranges over
the reals.  If instead we had taken time to range over the
natural numbers,
then would in fact get downward continuity.

We encourage the reader to check that $C\subG\varphi$, $\Ce\subG
\varphi$, and $\Cd\subG \varphi$ all satisfy the fixed point axiom and the
induction rule.  The fixed point axiom is a special case of the more
general fixed point axiom $\nu X. \varphi  \equiv \varphi
[\nu X.\varphi / X]$, while the induction rule is
a special case of the more general induction rule for fixed points:
from $\psi \supset \varphi[\psi/X]$
infer $\psi \supset \nu X . \varphi $.
The reader might also now wish to check that $C$
has the properties of S5, while $\Ce$ and $\Cd$
satisfy the positive introspection axiom and the necessitation rule.
Furthermore, for stable fact~$\varphi$ and complete-history
interpretations, they also satisfy the consequence closure axiom.
$\Ce$ and~$\Cd$ satisfy neither the knowledge axiom nor the negative
introspection axiom.
We remark that both notions satisfy weaker variants of the knowledge axiom:
$\Ce\varphi$ implies that~$\varphi$ holds at some point at most~$\eps$ time
units away from the current point, while~$\Cd\varphi$ implies that~$\varphi$
holds (at least) at some point during the run.
 
It is straightforward to extend the above framework to include
explicit individual clock times in order to define $C^T\subG\varphi$
(see \cite{NT} for more details).
Here, for example, it is the case that~$(\Ets(\varphi\wedge X))^\cM$
is downward continuous, and $\Ets$ distributes over conjunction;
hence~$\Cts$ will coincide with the appropriate infinite conjunction.
Similar treatments can be applied to many related variants of common knowledge
(see, for example, \cite{FH3,Mos,PT}).

\section*{Appendix B}

In this appendix we fill in the details of the proof that
common knowledge cannot be attained in practical systems
(Theorem~\ref{PracProp} in Section~\ref{AttSection}).
 
Our first step is to establish a general condition---namely,
that the initial point of a run is reachable from any later point---under
which
common knowledge can be neither gained nor lost.
We remark that Chandy and Misra have shown that in
the case of completely asynchronous, event-driven systems
where communication is not guaranteed,
common knowledge of any fact can be neither gained nor lost \cite{ChM}.
Since it is easy to see that, in such systems,
the initial point of a run is reachable
from all later points, our result
provides a generalization of that of \cite{ChM}.
 
\begin{proposition}%
\label{new}
Let $r\in R$ be a run in which the point $(r,0)$ is $G$-reachable
from $(r,t)$ in the graph corresponding to the complete-history
interpretation, and let $\I$ be a knowledge interpretation for $R$.
Then for all formulas $\varphi$ we have
$(\I,r,t) \sat C\subG\varphi$ iff $(\I,r,0) \sat C\subG\varphi$.
\end{proposition}
 
\begin{proof}
Fix a run $r$, time $t$, and formula $\varphi$.  Since $(r,0)$
is $G$-reachable from $(r,t)$ in the graph corresponding to the
complete-history
interpretation, there exist points $(\rz,\tz)$, $(\rone,\tone)$,
\ldots, $(r_k,t_k)$ such that $(r,t)=(\rz,\tz)$, $(r,0)=(r_k,t_k)$, and
for every $i<k$ there is a processor $j_i\in G$ that has the
same history at $(r_i,t_i)$ and at
$(r_{i+{\sts 1}},t_{i+{\sts 1}})$.
We can now prove by induction on $i$, using Lemma~\ref{Lemma1},
that $(\I,r,t) \sat C\subG\varphi$ iff $(\I,r_i,t_i) \sat C\subG\varphi$.
The result follows.  \qed
\end{proof}
 
We next provide a formal definition of systems with temporal imprecision,
and show that in such systems, the initial point of a run
is always reachable from later points.
A system $R$ has {\em temporal imprecision\/} if
$$ \begin{array}{l}
\forall r \in R \: \forall t \ge 0 \: \forall i \: \forall j \ne i \:
\exists \delta > 0 \: \forall \delta' \in [0,\delta)
\, \exists r' \, \forall t' < t\\
\ \ (h(p_i,r,t') = h(p_i,r',t'+\delta') \land
h(p_j,r,t')=h(p_j,r',t'))). \end{array}$$
Intuitively, this means that processors cannot perfectly
coordinate their notions
of time in a system with temporal imprecision.  One processor
might always be a little behind the others.

By {\em reachable} in the following lemma we mean reachable (in the sense
of Section~6) with respect
to the view function defined by the complete-history interpretation.
 
\begin{lemma}
\label{lemmanew}
If $R$ is a system with temporal imprecision, then for
all runs $r \in R$ and times $t$, the point $(r,0)$ is reachable from
$(r,t)$.
\end{lemma}
 
\begin{proof}
Let $R$ be a system with temporal imprecision and $(r,t)$ be a
point of $R$.  Suppose $t \ne 0$ (otherwise clearly $(r,0)$ is reachable
from $(r,t)$).  Let $\tz$ be the greatest lower bound of the set
$\{ t' : (r,t'') {\rm \ is \ reachable \ from \ } (r,t) \ {\rm for \
all \ } t'' \in [t',t] \}$.  We will show that $(r,\tz)$ is reachable
from $(r,t)$ and that $\tz = 0$.  Since $R$ is a system with temporal
imprecision, there exists a $\delta$
such that for all $\delta'$
with $0 < \delta' < \delta$, there exists a run $r'$ such that
for all $t' \le t$, we have $h(\pone,r,t') = h(\pone,r',t'+\delta')$
and $h(p_i,r,t') = h(p_i,r',t')$ for $i \ne 1$.  If $\delta' < t' \le t$,
it follows that $(r',t')$ is reachable from $(r,t')$ and
$(r,t'-\delta')$ is reachable from $(r',t')$.
By transitivity of reachability, we have that $(r,t'-\delta')$
is reachable from $(r,t')$, and by symmetry, that $(r,t')$ is
reachable from $(r,t'-\delta')$.  It now follows that
$(r,t-\delta')$ is reachable from $(r,t)$ for all
$\delta' < min(\delta,t)$.  Thus $\tz \le t - min(\delta,t)$.
Furthermore, if $\delta' < min(\delta, t)$, then we know that
$(r,\tz+\delta')$ is reachable from both $(r,\tz)$ and $(r,t)$.
It thus follows that $(r,\tz)$ is reachable from $(r,t)$.
Finally, if $\tz \ne 0$, then we know that $(r,\tz-\delta')$
is reachable from $(r,\tz)$ (and hence from $(r,t)$) for
all $\delta' < min (\tz,\delta)$.  But this contradicts our
choice of $\tz$.  Thus $\tz = 0$, and $(r,0)$ is reachable
from $(r,t)$.  \qed
\end{proof}
 
Theorem~\ref{PracProp}  now follows as
an immediate corollary to Lemma~\ref{lemmanew} and
Proposition~\ref{new}.
 
We conclude by showing that many practical systems do indeed
have temporal imprecision (although the $\delta$'s involved in some
cases might be very small).
Perhaps through statistical data, we can assume that for every
communication link $l$ there are known lower and upper bounds $L_l$
and $H_l$ respectively on the message delivery time for messages over
$l$.   We assume that the message delivery
time on the link $l$ is always in the open interval $(L_l, H_l)$.
(We take the interval to be open here since it seems reasonable to
suppose that if the system designer considers it possible that
a message will take time $T$ to be
delivered, then for some sufficiently small $\delta > 0$, he will also
consider it possible that the delivery time
is anywhere in the
interval $(T - \delta, T+ \delta)$;
in this we differ slightly from \cite{DHS,HMM}.)
We define $f_l$ to be a
{\em message delivery function for link $l$} if $f_l : {\bf N}
\rightarrow (L_l, H_l)$.
A run $r$ is {\em consistent} with $f_l$ if for all $n \in {\bf N}$,
$f_l(n)$ is the delivery time of the $n^{\rm th}$ message in $r$ on
link $l$.  A system $R$ has {\em bounded but uncertain message
delivery times} if for all links $l$ there exist bounds $L_l < H_l$
such that for all runs $r \in R$ and all message delivery functions
$f_l : {\bf N} \rightarrow (L_l, H_l)$, there exists a run $r'$
which is identical to $r$ except that message delivery time
over the link $l$ is defined by~$f_l$.  More formally,
$r'$ is consistent with $f_l$ and for all $i$, processor $p_i$ follows
the same protocol, wakes up at the same time (\ie $t_{init}(p_i,r) =
t_{init}(p_i,r')$), and has the same initial
state and the same clock readings in both $r$ and $r'$.

We say $R$ is a system with
{\em uncertain start times}
if there exists $\delta_0 > 0$ such that
 given a run $r \in R$, a processor
$p_i$, and $\delta$  with $0 < \delta < \delta_0$,
there is a run $r'$ which is identical to $r$ except that $p_i$ wakes up
$\delta$ earlier in $r'$ with its clock readings (if there are
clocks in the system) shifted back by $\delta$.  More formally,
for all $j \ne i$, processor $p_j$ follows the same protocol, wakes up
at the same time, and has the same initial state in both $r$ and $r'$.
Moreover, for all $k$, the delivery time for the
$k^{\rm th}$
message on link $l$ (if there is one) is the same in both $r$ and $r'$.
All processors other than $p_i$
have the same clock readings in both $r$ and $r'$.
Processor $p_i$ starts $\delta$ later
in $r'$ than $r$, although it has the same initial state in both runs,
and $\tau(p_i,r,t) = \tau(p_i,r',t + \delta)$.
 
For any practical system, it seems reasonable to assume that there
will be some (perhaps very small)
uncertainty in start times and, even if message delivery is
guaranteed within a bounded time, that there is some
uncertainty in message delivery time.
These assumptions are sufficient to guarantee temporal imprecision,
as the following result, whose proof is a slight modification
of a result proved in \cite{DHS}
on the tightness of clock synchronization achievable, shows:
 
\begin{proposition}
\label{Proposition55}
A system with bounded but uncertain message delivery times and
\markj%
uncertain start times has temporal imprecision.
\end{proposition}
 
\noindent{\bf Sketch of Proof:}\quad
Let $(r,t)$ be a point of the system, and let~$p_i$ be a processor.
Let $\delta_0$ be as in the definition of uncertain start times.
Since only a finite number of messages are received by time~$t$ in~$r$,
there is some $\delta>0$ such that the delivery times of
these messages are more than~$\delta$ greater
than the lower bound for the particular link they were sent over, and more
than~$\delta$ less than the upper bound.
Choose $\delta' < \min (\delta_0, \delta)$
and some processor $p_i$.  Let~$r'$ be a run in
which all processors $p_j\ne p_i$ start at the same time and in the
same initial state as in~$r$, have the same clock readings (if there
are clocks), and
all messages between such processors take exactly the same time as in~$r$.
In addition, processor~$p_i$ starts~$\delta'$ time units later in $r'$
than in~$r$,
messages to~$p_i$ take~$\delta'$ time units longer to be delivered,
while messages from $p_i$
are delivered~$\delta'$ time units faster than in~$r$,
and $p_i$'s clock readings (if there are clocks)
are shifted by $\delta'$.
Such a run $r'$ exists by our assumptions.  It is not hard to check
that run $r'$ has the property that for all times $t'\le t$,
all processors $p_j\ne p_i$ have exactly the same history at time~$t'$
in both~$r$ and~$r'$, while processor~$p_i$ has the same history at $(r,t')$
and at $(r',t'+\delta')$. Since $(r,t)$ and~$p_i$ were chosen arbitrarily, it
thus follows that the system has temporal imprecision.
\qed

\bibliographystyle{alpha}
\bibliography{z}
 
\end{document}